\documentstyle[12pt,aaspp4]{article}
\newcommand{\pld}{$\phi_{\rm LD}$}
\newcommand{\pud}{$\phi_{\rm UD}$}
\newcommand{\teff}{T$_{\rm eff}$}

\slugcomment{to appear in the March 2000 issue of The Astronomical Journal}
\begin{document}

\title{Synthetic Spectra and Color-Temperature Relations of M Giants}
\author{M. L. Houdashelt,\altaffilmark{1} R. A. Bell}
\affil{Department of Astronomy, University of Maryland, College Park, MD 20742-2421}
\authoremail{mlh@astro.umd.edu; roger@astro.umd.edu}

\author{A. V. Sweigart}
\affil{Code 681, NASA/Goddard Space Flight Center, Greenbelt, MD 20771}
\authoremail{sweigart@gsfc.nasa.gov}

\and

\author{R. F. Wing}
\affil{Department of Astronomy, Ohio State University, Columbus, OH 43210}
\authoremail{wing@astronomy.ohio-state.edu}

\altaffiltext{1}{Current address: Department of Physics \& Astronomy, Johns Hopkins University, 3400 North Charles Street, Baltimore, MD 21218}

\begin{abstract}

As part of a project to model the integrated spectra and colors of elliptical
galaxies through evolutionary synthesis, we have refined our synthetic spectrum
calculations of M~giants.
After critically assessing three effective temperature scales for
M~giants, we adopted the relation of Dyck et al. (1996)
for our models.  Using empirical spectra of field M~giants as a guide, we then
calculated MARCS stellar atmosphere models and SSG synthetic spectra of these
cool stars, adjusting the band absorption oscillator strengths of the TiO bands
to better reproduce the observational data.  The resulting synthetic spectra
are found to be in very good agreement with the K-band spectra of stars
of the appropriate spectral type taken from Kleinmann \& Hall (1986) as well.
Spectral types estimated from the strengths of the TiO bands
and the depth of the bandhead of CO near 2.3~$\mu$m quantitatively confirm
that the synthetic spectra are good representations of those of field
M~giants.  The broad-band colors of the models match the field
relations of K~and early-M giants very well; for late-M giants, differences
between the field-star and synthetic colors are probably caused by the omission
of spectral lines of VO and H$_2$O in the spectrum synthesis calculations.
Here, we present four grids of K-band bolometric corrections and colors --
Johnson U--V and B--V; Cousins V--R and V--I; Johnson-Glass V--K, J--K and
H--K; and CIT/CTIO V--K, J--K, H--K and CO -- for models having
3000~K~$\leq$~\teff~$\leq$~4000~K and --0.5~$\leq$~log~g~$\leq$~1.5.  These
grids, which have [Fe/H]~=~+0.25, 0.0, --0.5 and --1.0, extend and supplement
the color-temperature relations of hotter stars presented in a companion paper.

\end{abstract}


\section{Introduction}

M~giants are important contributors to the integrated light of many stellar
aggregates, such as early-type galaxies and galactic bulges, even though they
generally comprise a very small fraction of the stellar
mass of these systems.  In fact, recent integrated light models of the
Galactic bulge of the Milky Way (\cite{houdy}) indicate that M~giants
contribute over half of the K-band flux there and thus also dictate the strength
of spectral features such as the 2.3 $\mu$m CO band.
However, the relative importance of these cool stars in integrated light
is quite dependent upon the wavelength region under consideration and the
other stellar populations present.  For example, at optical wavelengths, where
the Galactic-bulge M giants are much fainter than they are in the near-infrared,
these stars produce only about 10--20\% of the bulge's continuous flux but are
entirely responsible for the broad absorption bands of TiO seen in this part of
its spectral energy distribution.  Consequently, population models of
galaxies should strive to include realistic representations of M~giants.

Unfortunately, M~stars have proven especially difficult to model accurately
because they are so cool and thus have a wealth of molecules in their
stellar atmospheres.  This means that a
variety of phenomena which can be ignored in the modelling of
hotter stars are important in the atmospheric structure of M~stars.
The most critical of these are the molecular opacities, which depend
not only on the accuracy of the (sometimes nonexistent) laboratory data
but also on the way in which the opacities are calculated and
represented in the models.  Simple mean opacities, opacity distribution
functions and opacity sampling have each been used in the modelling
of cool star atmospheres.  For example, plane-parallel models of M~giants
have been calculated by \cite{brett} using straight mean opacities and by
\cite{jorg94} using opacity sampling.

Other factors influencing cool star models include sphericity effects and
variability.  In spherical models, an additional parameter is introduced --
the extension of the atmosphere, d.  It is defined by the relation,
d~=~r/R~--~1, where R is the stellar radius, typically defined to be
the radius at which the Rosseland optical depth is equal to unity, and
r is the radius at which the Rosseland optical depth has a value of 10$^{-5}$.
Static, spherical models of M~giants have been constructed by \cite{bbsw1},
\cite{st84}, \cite {scholz} and \cite{plez92a}.  All but the latter used
straight mean opacities in their models; \cite{plez92a} incorporated more
recent opacity data and used opacity sampling.
Dynamic, spherical models, representing Mira variables, have
been studied by \cite{bbsw2} and \cite{ap98}.

Extension and sphericity are generally
important for two reasons.  First, if log~g varies significantly between the
base of the stellar atmosphere and its outer layers, it may be necessary to
include the radial dependence of gravity explicitly in the stellar atmosphere
model.  Second, extension results in a higher photon escape probability, and
the corresponding dilution of the stellar flux produces a cooling of the
outer layers of the atmosphere.  Due to the
temperature sensitivity of molecule formation, this cooling in turn enhances
the formation of certain molecules and thus their partial pressures.
In fact, extension can be so important in cool stars that \cite{sw82} and
\cite {scholz} have suggested a 3-dimensional classification scheme for
M~giants in which extension serves as the third
parameter (in addition to \teff\ and log~g); they propose that the
extension of an M~giant can be estimated observationally from the depths of
specific TiO bands at a given effective temperature and surface gravity.
However, \cite{plez92a} has found that a good representation of the opacities
is more important than sphericity effects for models having
log~g~$\gtrsim$~0.0 and masses of order 1~M$_{\sun}$ or more.  Thus, sphericity
and variability significantly affect only the coolest of M~giant models.

We have recently begun an evolutionary synthesis program to produce synthetic
spectra of early-type galaxies.  The foundations of this work are stellar
atmosphere models and synthetic spectra calculated with updated versions of
the MARCS (\cite{gben}, \cite{begn}) and SSG (\cite{bg78}; \cite{gb79}; Bell
\& Gustafsson 1989; hereafter \cite{bg89})
computer codes, respectively.  However, because we employ versions of these
codes which do not account for sphericity and other factors which affect
the stellar atmospheres of M~giants, we have exerted a
considerable effort to fine-tune our models to compute more representative
synthetic spectra of these cool stars and also to establish an effective
temperature scale for M~giants based upon recent angular diameter measurements.
In a companion paper, Houdashelt et al. (2000; hereafter \cite{hbs1}), we
describe many of the recent improvements in these codes and
present new color-temperature relations for stars as cool as spectral type~K5.

In the present paper, we discuss our improved models of M~giants and compare
the resulting synthetic spectra and colors to observational data.
In Section 2, we briefly describe the MARCS/SSG models and calculations,
examine the possible shortcomings of using these models to represent M~giants
and present our strategy for testing and refining the synthetic spectrum
calculations to compensate for these shortcomings.  We compare three
effective temperature scales for cool stars in Section 3 and determine which
best represents the field M~giants.  Section 4 discusses our treatment of TiO
absorption in the synthetic spectra and compares our results to observed
spectra of field M~giants.  We also show the good agreement between
the computed and observed CO band strengths and compare the broad-band colors
measured from the synthetic spectra to photometry of field stars.
Section 5 summarizes the major conclusions of this work.

\section{Basic Details of the M Giant Models}

The models of M~giants presented in this paper have been constructed in
exactly the same manner as those of the hotter stars described in \cite{hbs1}.
We provide here only a brief description of the MARCS model atmospheres and
the SSG synthetic spectrum calculations,
emphasizing those factors which are most relevant to calculating models of
cool stars.  We refer the reader to \cite{hbs1} for further details.

\subsection{Calculating the Model Atmospheres and Synthetic Spectra}

The version of the {\sc MARCS} stellar atmosphere code
used to construct the model atmospheres of the M~giants
produces a flux-constant, chemically-homogeneous, plane-parallel model
atmosphere calculated under the assumptions of 
hydrostatic equilibrium and LTE.  It incorporates opacity distribution
functions (ODFs) to represent the opacity due to atomic and molecular lines
as a function of wavelength.

The SSG spectral synthesis code combines the MARCS model atmosphere and
spectral line lists to compute a synthetic spectrum.  The primary spectral
line list which we use is the updated
version of the Bell ``N'' list (\cite{bpt94}) described
in \cite{hbs1}.  In addition, optional line lists for TiO and H$_2$O can
be included in the calculations.  We have incorporated the TiO line list
in our M giant models, and we describe it more fully below.  As described
later in this paper (see Section 4.2.3), both of the H$_2$O line lists which
we tested were found to be unsatisfactory, so no water lines are included in
the models presented here.

The TiO spectral line list includes lines from the $\alpha$ (C$^3\Delta$--X$^3\Delta$),
$\beta$ (c$^1\Phi$--a$^1\Delta$), $\gamma$ (A$^3\Phi$--X$^3\Delta$),
$\gamma^{\prime}$ (B$^3\Pi$--X$^3\Delta$), $\delta$ (b$^1\Pi$--a$^1\Delta$),
$\phi$ (b$^1\Pi$--d$^1\Sigma$), and $\epsilon$ (E$^3\Pi$--X$^3\Delta$) systems.
In addition to the lines of $^{48}$TiO, the spectral line lists include lines
of $^{46}$TiO, $^{47}$TiO, $^{49}$TiO and $^{50}$TiO as well.
Wavelengths and gf values for lines in the $\epsilon$ system were kindly
provided by \cite{plez96}.  For lines of the other systems, wavelengths were
calculated using molecular constants taken from \cite{phil73}, and the
H\"{o}nl-London factors were obtained from formulae in \cite{kovacs}.
The Franck-Condon factors were taken from \cite{bdbh} for the $\alpha$, 
$\gamma$, $\gamma^{\prime}$ and $\phi$ systems and computed for the $\beta$ and
$\delta$ systems using the code described by \cite{bbu}.  The initial f$_{00}$
values
for the $\alpha$, $\beta$, $\gamma$ and $\gamma^{\prime}$ systems came from
\cite{hnc95}, while those for the $\delta$ and $\phi$ bands were assumed to
be 0.0190 and 0.0210, respectively.  Improved values were found
empirically by comparing observed and synthetic spectra of field M~giants
as described in Section 4.1.

All of the stellar atmosphere models and synthetic spectra discussed in this
paper have been constructed using solar abundance ratios for all of the
elements except carbon and nitrogen.  \cite{hbs1} discusses the evidence
indicating that stars which are more evolved than the ``bump'' in the
red-giant-branch luminosity function have had CNO-processed material mixed into
their atmospheres.  Consequently, we have used [C/Fe]~=~--0.2,
[N/H]~=~+0.4 and $^{12}$C/$^{13}$C~=~14 for our M~giant models, in
accordance with the abundance ratios measured in field M~giants (\cite{sl90}).

As in \cite{hbs1}, the synthetic spectra were calculated at 0.1~\AA\ resolution
and in two pieces, optical and infrared (IR).  The optical portion of the
spectrum covers wavelengths from 3000--12000~\AA, and the IR section
extends from 1.0--5.1~$\mu$m (the overlap is required for calculating J-band
magnitudes).  In addition, the microturbulent velocity, $\xi$, used to 
calculate the synthetic spectrum of a given
star was derived from its surface gravity using the field-star relation,
$\xi$~=~2.22~--~0.322~log~g (\cite{gcc96} 1996).

\subsection{Possible Deficiencies in the Models}

There are two possible drawbacks to using our version of MARCS model
atmospheres to represent M~giants.  First,
being plane-parallel models, they do not account for the affects of sphericity
and extension.  Second, the ODFs which we employ do not include the molecular
opacities of TiO, VO and H$_2$O, which are among the strongest
opacity sources in M~stars.  In addition, the spectral line list used by
SSG does not include lines of VO.  Still, there are good reasons to believe that
the models, especially those of the hotter M~giants, will not be too greatly
in error due to these factors.

\cite{jorg94} has
shown that the inclusion of TiO in plane-parallel model atmospheres mainly
serves to heat the surface layers of the models, with the extent of the
heating diminishing as \teff\ decreases.  This heating inhibits the formation
of H$_2$O, which thus does not become an important source of opacity
until mid-to-late-M~types.  The main effect of H$_2$O opacity in his models
is to cause an
expansion of the atmosphere in cooler stars (\teff~$\lesssim$~3000~K).  VO
appears to have little influence on the atmospheric structure but does affect
the spectra of M~giants, producing a series of broad absorption bands between
about 0.7 and 2.2 $\mu$m in the spectra of giants later than spectral type M5
(Brett 1990; Plez 1998).

From a physical standpoint, the basic differences between spherical model
atmospheres and their
plane-parallel counterparts appear to be 1) the spherical models extend to
lower gas pressures and temperatures, and 2) as extension increases, the
temperature at a given optical depth decreases, while the gas pressure at a
given temperature increases (see e.g., Scholz \& Tsuji 1984; Scholz 1985).
\cite{scholz} and \cite{bbsw1} have found that
extension is relatively unimportant in early-type M~giants but dramatically
increases with decreasing effective temperature for \teff~$\lesssim$~3500~K,
mainly due to a substantial increase in the H$_2$O opacity.  In addition,
they find that extension influences only the uppermost layers of the atmosphere,
so that molecular species such as TiO, VO and H$_2$O are affected, but CN
and CO, which mostly form deeper in the atmosphere, are found to be relatively
insensitive to extension.
\cite{plez92a} also find that extension affects TiO formation mainly because it
forms in the upper layers of the atmosphere; extension is the greatest for 
their models near 3200~K due to the saturation of H$_2$O bands.

Thus, while these factors (plane-parallel model atmospheres, missing opacity
in the ODFs) are potential hindrances to successful
modelling of cool giants, we expect them to have relatively minor effects on
our results.  Since they affect only the surface layers of the models, the
continuum in our models should accurately represent the effective temperature.
On the other hand, we would be surprised to find that the absorption bands
of the molecules formed in the outer parts of the atmosphere matched those
observed in stars of the corresponding \teff.  The TiO bands, which are
extremely temperature-sensitive, should be the most important in this regard,
since VO and H$_2$O bands are observed in the spectra of only the coolest
M~giants (spectral type M5 and later) and/or those of very low gravity;
the omission of the latter two molecules in the ODFs probably has a minimal
influence on the stellar atmosphere models of most M giants.  Thus, we have
good reason to believe that we can overcome the deficiencies inherent in our
ODFs and plane-parallel models by modifying our treatment of TiO.

\subsection{How Do We Test Our Models?}

Quite often, stellar models are evaluated by comparing the colors of a grid of
solar-metallicity
models to observed color-color or color-temperature relations of field stars.
If the field star relations fall within the domain of the grid colors at a
given effective temperature or color, then the models are usually considered
to be satisfactory.  The fact that this approach is only truly appropriate for
colors which are reasonable representations of the continuum slope and are
relatively insensitive to gravity is often ignored.  While this may be
reasonable for many of the colors of hotter stars, such effects must not be
neglected when modelling M~giants, since their spectra are
dominated by molecular absorption bands and gravity-sensitive features.
It is possible that, through a
fortuitous but incorrect combination of log~g, [Fe/H] and perhaps
microturbulent velocity, a synthetic spectrum can be calculated 
which has the colors of a field M~giant of a given effective temperature
but proves to be a poor match to the finer details of its spectral energy
distribution.  Unfortunately,
without {\it a priori} knowledge of the temperatures and gravities of the
field stars, a better way to test the model colors is not obvious.

Given the aforementioned uncertainties in our models and in dealing with cool
stars in general, we do not expect to be able to produce perfect synthetic
spectra of M~giants.  However, we need to be able to evaluate the ``quality''
of our models and refine them as necessary.  We have chosen to do this
by comparing our synthetic spectra
to observed spectra of field M~giants in as much detail as possible.
Ideally, this would utilize a good set of empirical spectra of stars of
known \teff, gravity and metallicity; to the best of our knowledge,
such a set of data does not exist for M~giants.  Instead, we have chosen to use
the ``intrinsic'' spectral sequence of field M~giants presented by
Fluks et al. (1994; hereafter \cite{fluks}) to test and improve our synthetic
spectrum calculations.  However, to use this data for such a purpose, we must
first assign effective temperatures and surface gravities to the stars
represented by their spectral sequence.

In the MK spectral classification system, temperature classes of M~giants are
based primarily upon the strengths of the TiO bands (see \cite{km76}).  While
this allows an MK spectral type to be assigned to any spectrum
containing TiO bands, it also makes the
classification of these cool stars metallicity-dependent, since the
TiO band strengths depend upon the abundances of titanium and oxygen (we adopt
logarithmic abundances of 4.78 and 8.87 dex for Ti and O, respectively, on a
scale where H~=~12.0 dex).  In other words, an
M2~giant with solar abundances will not have the same effective temperature as
a metal-poor M2~giant or an M2~giant with
non-solar Ti/Fe ratios, such as stars in the Galactic bulge (\cite{mr94}).
Because the TiO bands are also sensitive to gravity (see Bessell et al.  1989a)
and extension (for the coolest and lowest gravity M~stars), there may be a
more complex relation between spectral type and effective temperature for
M~giants than for hotter stars.
Nevertheless, the primary factor affecting the TiO bands is \teff, and
the crucial step in calculating
realistic models of M~giants is to reproduce the specific relationship between
the depths of the TiO bands (i.e., spectral type) and effective temperature
which is observed in field M~giants.

Thus, the first step in our modelling is to determine the spectral type,
\teff\ relation (hereafter STT relation) which holds for the field~M giants;
we do
this in the subsequent section of this paper.  We then assign a surface gravity
to each M~giant using a relation between log~g and \teff\ derived from the
isochrones and stellar evolutionary tracks produced as part of our
evolutionary synthesis program (see
\cite{hbs1} and \cite{hbs2} for details).  As expected from the
discussion of our cool star models, the strengths of some of the TiO bands
in the initial synthetic spectra failed to match the observed
spectra, and we were forced to adjust the f$_{00}$ values of the individual
TiO bands to improve the overall agreement between the empirical and synthetic
spectra.

\section{The M Giant Temperature Scale}

As discussed in \cite{hbs1}, the most direct way to determine the effective
temperature of a star is through measurement of its angular diameter, $\phi$,
and its apparent bolometric flux, f$_{\rm bol}$.
These parameters can be related to effective temperature through the relation,
\begin{equation}
{\rm T}_{\rm eff} \propto \left( {\rm f}_{\rm bol} \over \phi^2 \right)^{0.25}\ .
\end{equation}
Of course, angular diameters can be determined for only the most nearby stars
and are typically measured through either lunar occultations or interferometry
(e.g., speckle, intensity, Michelson).  Usually, the diameter
measured by these methods is that for a uniform disk, denoted
\pud, which must then be adjusted to give the limb-darkened angular
diameter, \pld, before computing the effective temperature.
The limb-darkening correction
is generally derived from stellar atmosphere models and is currently one of
the greatest uncertainties in estimating the effective temperature, since
this correction is sensitive to stellar parameters (such as T$_{\rm eff}$
itself) and the wavelength at which \pud\ is determined.

As our main purpose in modelling M~giants is to use their synthetic spectra
for evolutionary synthesis of galaxies, the temperature scale that we choose
must be consistent with that used in \cite{hbs1} for the hotter stars, the
STT relation of \cite{bg89}.  Since \cite{bg89} only
derived effective temperatures for G~and K~stars, we must find a
complementary relation for cooler stars.  Below, we
examine three different temperature relations for M0--M7 giants, each
based upon angular diameter measurements -- the
relation given by Dyck et al. (1996; hereafter \cite{dbbr}), that of
Di Benedetto \& Rabbia (1987; hereafter \cite{dib87}), and a temperature scale
which we have derived from angular diameters measured by \cite{moz91} and
Mozurkewich (1997), which we will hereafter collectively refer to as
\cite{moz97}.  \cite{perrin} have recently
estimated \teff\ for stars even later than spectral type~M7, but we have not
included their results because even the most metal-rich isochrones used in our
evolutionary synthesis models do not extend to such cool effective temperatures.

\subsection{Angular Diameter Measurements}

\cite{dbbr} combined interferometric angular diameters of 34~stars measured
with the Infrared Optical Telescope Array at 2.2~$\mu$m and
occultation diameters from \cite{rjww} to derive a relation between effective
temperature and spectral type
for K~and M~giants.  Their temperatures were determined using the
limb-darkening correction \pld~=~1.022~\pud, which they
derived from the stellar atmosphere models of \cite{st87}.
\cite{dbbr} estimated the uncertainty in their effective
temperature at a given spectral type to be approximately 95~K.  They also
concluded that M~supergiants were systematically cooler than M~giants
of the same spectral type.

Di~Benedetto (1993; hereafter \cite{dib93}) tabulated angular diameters of
21~stars, primarily taken from the work of \cite{dib87} and \cite{dibf}.
These angular diameters were also measured at 2.2~$\mu$m using Michelson
interferometry, but the limb-darkening corrections were derived from 
the models of \cite{man79}; they used 1.026~$\leq$~\pld/\pud~$\leq$~1.036, with
an average of 1.035 for stars later than spectral type K5.  \cite{dbbr} noted
that their uniform-disk angular diameters agreed well with those of \cite{dib87}
for \pud\ $\leq$~10~mas but were systematically smaller (by about 10\%
on average) than the measurements of \cite{dib87} for larger stars.

\cite{moz97} has presented uniform-disk angular
diameters measured with the Mark III Interferometer at 8000~\AA.  To convert
these measurements to limb-darkened diameters, we used the
giant-star data presented in Table~3 of \cite{moz91} to derive the relation,
\pld/\pud~=~1.078~+~0.002139~SP, where SP is the M~spectral
class of the star (e.g., SP~=~0 for an M0 star, SP~=~--1 for a K5~star).
\cite{hbs1} gives further details of the derivation of this relation.

Since \cite{bg89} were required to estimate apparent bolometric fluxes to
compute effective temperatures using the infrared-flux method, they were also
able to predict angular diameters of the stars they examined.  Ideally,
we would like to compare their predicted diameters to those measured by
the other groups.  Unfortunately, the samples of \cite{dbbr} and \cite{dib93}
have very little overlap with \cite{bg89}: two and four stars,
respectively.  Thus, to examine the compatibility of the four sets of angular
diameters, we compare the \cite{bg89}, \cite{dbbr} and \cite{dib93}
data to the measurements of \cite{moz97}.

\begin{figure}[p]
\epsfxsize=6.5in
\vspace*{-1.9in}
\epsfbox{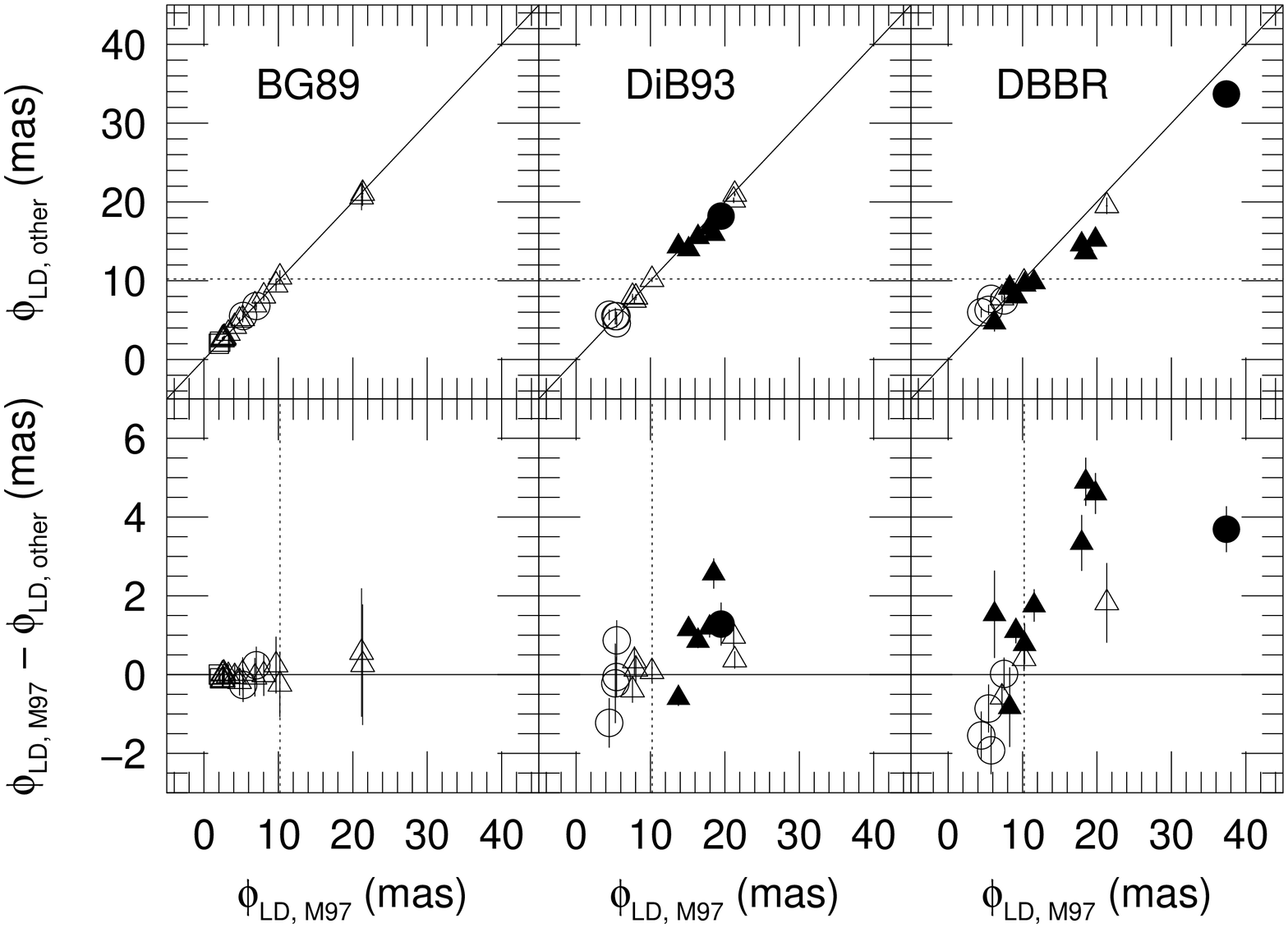}
\vspace*{-0.3in}
\caption{Comparison of limb-darkened angular diameter
measurements (\pld) of G--M stars.  The upper panels show direct comparisons
of the \pld\ measurements; the lower panels show the \pld\ differences.
From left to right, the following comparisons are made: Bell \& Gustafsson
(1989; BG89) vs. Mozurkewich et al. (1991) and Mozurkewich (1997; M97
collectively), Di~Benedetto (1993; DiB93) vs. M97, and Dyck et al.
(1996; DBBR) vs. M97.  The error bars for the M97,
DiB93 and DBBR angular diameters come directly from the respective
references; those for the BG89 diameters were calculated as described
in the text.  Solid lines show equality of the diameters.  Dotted lines appear
at 10.22~mas; DBBR noted systematic differences between their diameters
and those of Di~Benedetto \& Rabbia (1987) for stars larger than this.
Supergiants, giants and subgiants are shown as circles, triangles and squares,
respectively; open symbols are G~and K~stars, and filled symbols are M stars.}
\label{angdiamcomp}
\vspace*{0.5in}
\end{figure}

The upper panels of Figure~\ref{angdiamcomp} show direct comparisons of these
angular diameters, and the corresponding bottom panels illustrate
the differences between the diameters plotted in the upper panels.  The error
bars shown for the \cite{bg89} angular diameters have been calculated using
1.316~$\times$~10$^7$ as the constant of proportionality in equation~1
(\cite{dbbr}) and assuming a 4\% uncertainty in f$_{\rm bol}$ and an uncertainty
of 150~K in \teff\ (see \cite{bg89}).  The other error bars have been taken
directly from the respective references.  In each
of the panels of Figure~\ref{angdiamcomp}, the
solid line represents equality of the two sets of measurements compared there;
the dotted line shows the angular diameter above which \cite{dbbr} noted that
their diameters differed systematically from those of \cite{dib87}.
Squares, triangles and circles have been used to represent subgiants, giants
and bright giants, and supergiants, respectively; filled symbols show M~stars,
and open symbols are G~and K~stars.  A quantitative comparison of the
\pld\ measurements is given in Table~\ref{angdiamtable}.

Several conclusions can be drawn from Figure~\ref{angdiamcomp} and
Table~\ref{angdiamtable}.  First,
the \cite{moz97} and \cite{bg89} angular diameters are remarkably similar, a
point already emphasized in \cite{hbs1}.  Second, while all three sets of 
measurements appear to be consistent for \pld\ $\leq$~10.22~mas
(\pud\ $\leq$~10~mas), the \cite{moz97} angular diameters are systematically
greater than the others for stars larger than this.  Third, the differences
between the four sets of angular diameters are dominated
by the M~star measurements, probably because the stars with \pud\ $\geq$~10~mas
tend to be M~stars; in fact, the \pld\ values for the G~and K~stars
do not show systematic differences.

From the comparisons displayed in Figure~\ref{angdiamcomp} and summarized in
Table~\ref{angdiamtable}, we
conclude that the \cite{moz97}, \cite{bg89}, \cite{dbbr} and \cite{dib93}
angular diameters are in sufficient agreement for G~and K~stars to infer
that effective temperatures based upon any of the other three group's
measurements would be consistent with the \cite{bg89} temperature
scale for these stars, although there is certainly less scatter in the
comparison with \cite{moz97}.  However, the same conclusion cannot be drawn for
the M~stars, and we must evaluate the resulting STT relations
individually to determine which is the most suitable for producing accurate
synthetic spectra of M~giants.

\newpage

\subsection{Comparing Effective Temperature Scales}

\cite{dbbr} and \cite{dib87} have derived very similar STT relations, even
though some systematic differences exist in their angular diameter measurements
for K~and M~giants, evidently because systematic differences in their
estimates of f$_{\rm bol}$ largely offset the \pld\ differences.
To derive analogous STT relations based upon \cite{moz97}'s angular diameters,
we have simply adopted the bolometric fluxes of \cite{dbbr} and \cite{dib87}
to calculate ``\cite{moz97}'' effective temperatures for the stars that each
group had in common with \cite{moz97}.

\begin{figure}[p]
\epsfxsize=6.5in
\vspace*{-0.5in}
\epsfbox{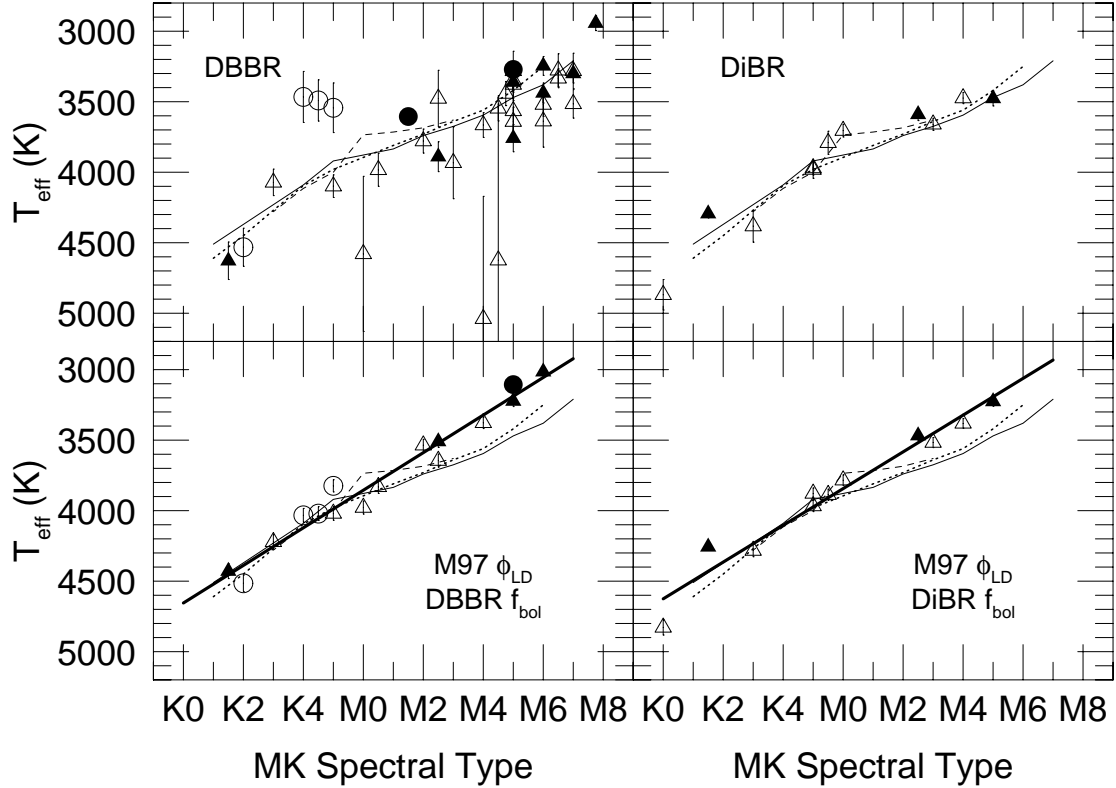}
\vspace*{-0.3in}
\caption{Spectral type, effective temperature relations for
K--M giants.  The upper, left-hand and upper, right-hand panels show the
data (and error bars) reported by Dyck et al. (1996; DBBR) and
Di~Benedetto \& Rabbia (1987; DiBR), respectively.  The lower panels
show the effective temperatures which result when the angular diameters of
Mozurkewich et al. (1991) and Mozurkewich (1997; M97 collectively) are
substituted for those of DBBR and
DiBR in the \teff\ calculations of the stars in the
respective upper panels which M97 observed.  The calculation of the
temperature errors for these data is described in the text.
Supergiants are represented by circles,
and giants are triangles.  Open symbols are stars with \pud\ $<$~10~mas;
filled symbols are larger stars.  The solid, dotted and dashed lines in
each panel show the relations tabulated by DBBR, Ridgway et al. (1980) and
DiBR, respectively.  The bold lines in the lower panels are linear,
least-squares fits to the data shown there.}
\label{typetemp}
\end{figure}

The effects of using the various \pld\ and f$_{\rm bol}$ measurements when 
calculating effective temperatures are
shown in the four panels of Figure~\ref{typetemp}.  The upper section of
this figure shows the data reported by \cite{dbbr} and
\cite{dib87} in the left-hand and right-hand panels, respectively; the
lower panels show the effective temperatures which result when the angular
diameters of \cite{moz97} are substituted for those used in the corresponding
upper panels for the stars in common.  As in Figure~\ref{angdiamcomp}, triangles
represent giants and circles are supergiants, but
the filled symbols here are stars having \pud\ $>$~10~mas, while the open
symbols are those with smaller angular diameters.  The temperature errors
shown in the upper panels of the figure are those quoted by \cite{dbbr} and
\cite{dib87}; those in the lower panels have been derived from equation~1,
using 1.316~$\times$~10$^7$ as the constant of proportionality (\cite{dbbr})
and adopting the \pud\ uncertainties of \cite{moz97} and the flux uncertainties
of either \cite{dib87} or \cite{dbbr}, as appropriate.  The dotted line in each
panel of Figure~\ref{typetemp} is the STT relation quoted by \cite{rjww} and
is based upon angular diameters measured from lunar occultations (not shown).
The solid line is \cite{dbbr}'s relation, which incorporates the \cite{rjww}
measurements in addition to those plotted in the upper, left-hand panel of
Figure~\ref{typetemp}.  The dashed
line is the STT relation of \cite{dib87}.  The bold lines in the lower panels
of Figure~\ref{typetemp} are linear, least-squares fits to the
data plotted in each.

From Figure~\ref{typetemp}, it is clear that the STT relations of \cite{dbbr}
and \cite{dib87}
are very similar; the greatest difference occurs near spectral type M0.  It is
also true that the STT relations derived from the \cite{moz97} angular
diameters (the bold, solid lines in the lower panels of Figure~\ref{typetemp})
are nearly identical, regardless of whether the \cite{dbbr}
or \cite{dib87} bolometric fluxes are used; in fact, the resulting relations
differ by less than 30~K at all spectral types from K0~to~M7.  Thus, for the
remainder of this paper, we will no longer discuss the STT relation
of \cite{dib87} or the relation derived from the \cite{dib87} bolometric fluxes
using the \cite{moz97} angular diameters; we will instead concentrate upon
the \cite{dbbr} data because it spans a broader range of spectral types.

As expected from the comparison of the \cite{moz97} and \cite{dbbr} angular
diameter measurements, the STT relations which result from these
two data sets agree nicely for the K~giants but differ systematically in the
M~giant regime.  To quantify this, we tabulate these two \teff\ scales and our
estimates for the surface gravities of field M~giants in Table~\ref{tefftable};
the calculation of these log~g values is described below.
We will hereafter refer to these two STT relations as the \cite{dbbr} and
\cite{moz97} effective temperature scales.

Note that all of the spectral types plotted in Figure~\ref{typetemp} are those
adopted by \cite{dbbr} and \cite{dib93} (which are identical to
\cite{dib87}'s) and are presumably MK types, since
they were generally taken from the work of Keenan and collaborators.
We have confirmed this for the 29 stars observed by \cite{dbbr} for which
spectral types have also been measured using the 8-color photometric system of
\cite{wing71}; Wing's method (see Section 4.2.1) provides a quantitative way
to measure spectral types of M giants on the MK system.  The average difference
in spectral types, in the sense Wing -- \cite{dbbr}, is 0.29 ($\pm$0.42)
subtypes.  If we use Wing's spectral types, when available, to revise the 
\cite{moz97} STT relation in the bottom, left-hand panel of
Figure~\ref{typetemp}, it only differs from that adopted (Table~\ref{tefftable})
by +7~K at spectral type K1, --11~K at type M0 and --35~K at type M7.
Thus, \cite{moz97}'s angular diameter measurements truly infer a different
relation between \teff\ and MK spectral type than that derived by \cite{dbbr}.
Since it is not clear which temperature scale is the best
to use for constructing synthetic spectra of M~giants, we have experimented
with each and discuss the results in the following sections.

\subsection{Testing the Effective Temperature Scales}

To determine which of these two STT~relations is best suited for M~giant
models, we have calculated model atmospheres and synthetic spectra (omitting
spectral lines of TiO) for M0~through M7~giants on each \teff\ scale given in
Table~\ref{tefftable}.  We assigned surface gravities to
these models by consulting the solar-metallicity isochrones and evolutionary
tracks used in our evolutionary synthesis program (\cite{hbs2}).
Specifically, we took log~g values at 100~K intervals between 3200~K and
4000~K from our 4~Gyr isochrone (3500--4000~K),
8~Gyr isochrone (3400~K), 16~Gyr isochrone (3300~K) and 0.7~M$_{\odot}$
evolutionary track (3200~K) and fit a quadratic relation to this data; this
relation was used to derive the surface gravities listed in
Table~\ref{tefftable}.

\begin{figure}[p]
\epsfxsize=6.0in
\vspace*{-0.9in}
\hspace*{0.25in}
\epsfbox{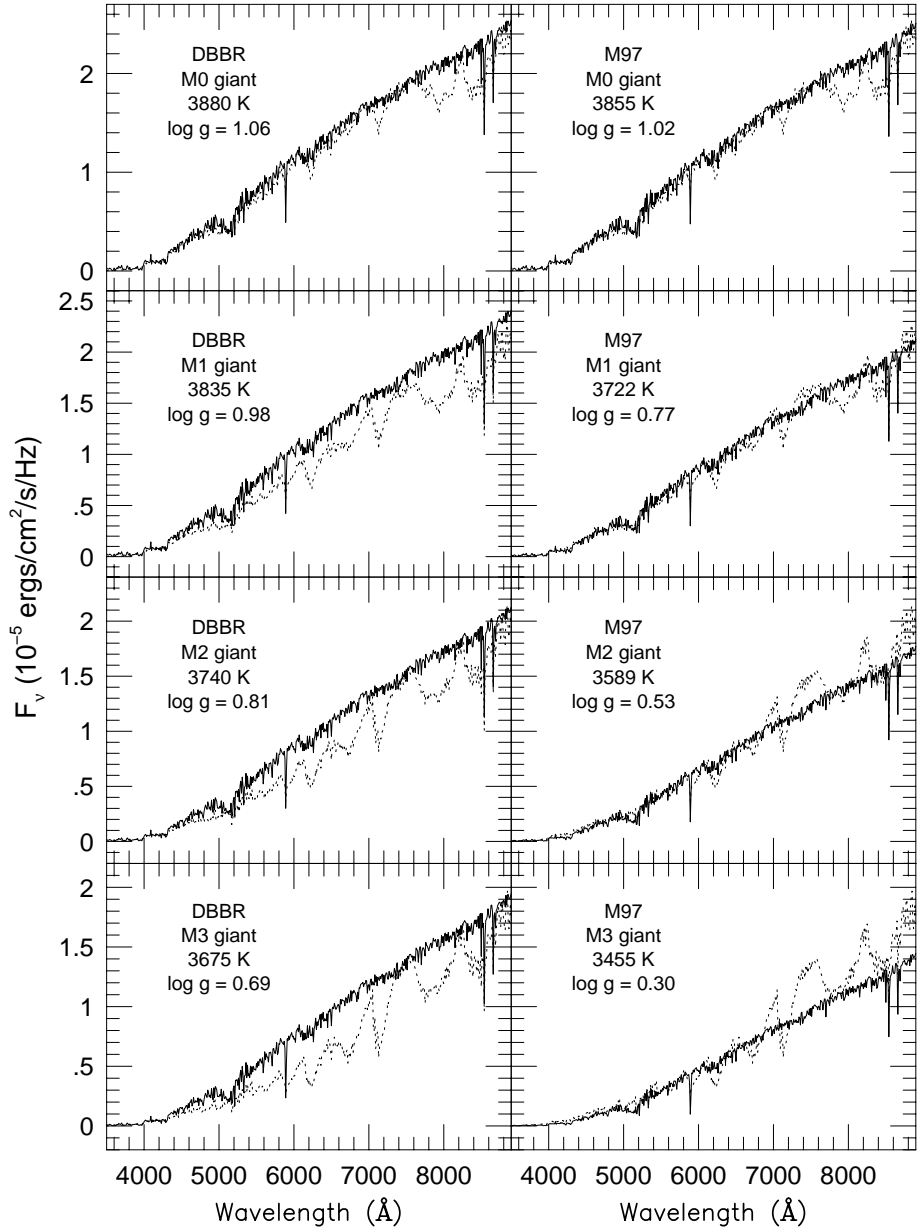}
\vspace*{-0.7in}
\caption{Comparisons of field-star spectra of M0--M3 giants
and our synthetic spectra which omit spectral lines of TiO.
The solid lines are the solar-metallicity synthetic spectra, and the dotted
lines are the ``intrinsic'' MK spectra of field M~giants from Fluks et al.
(1994).  The left-hand panels show synthetic spectra calculated assuming the
\teff, spectral type relation of Dyck et al. (1996; DBBR), and the
synthetic spectra in the right-hand panels result from the relation derived here
using the angular diameter measurements of Mozurkewich et al. (1991) and
Mozurkewich (1997; collectively M97).  The Fluks et al. spectra are the
same in adjacent left-hand and right-hand panels.  All panels are labeled
with the temperature relation adopted, the spectral type of the Fluks et al.
spectrum and the \teff\ and log~g values used in the corresponding models.}
\label{tempcomp1}
\end{figure}

\begin{figure}[p]
\epsfxsize=6.0in
\vspace*{-0.3in}
\hspace*{0.25in}
\epsfbox{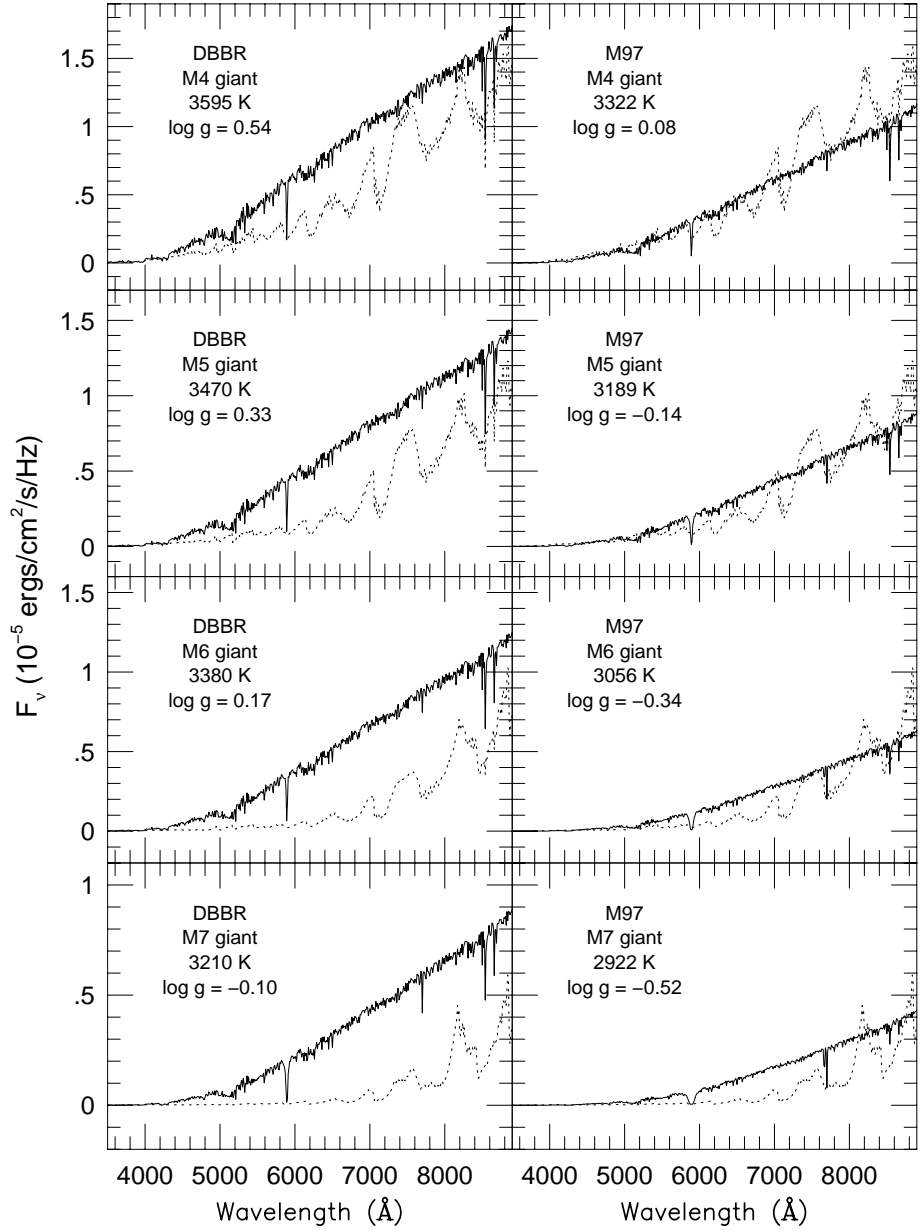}
\vspace*{-0.5in}
\caption{Comparisons of field-star spectra of M4--M7 giants
and our synthetic spectra which omit spectral lines of TiO.  See the caption
to Figure~3 for further details.}
\label{tempcomp2}
\end{figure}

We compare our synthetic spectra to the ``intrinsic'' MK spectra of field
M~giants presented by \cite{fluks} in Figures~\ref{tempcomp1}
and~\ref{tempcomp2}.  In these figures, the
synthetic spectra are shown as solid lines and the \cite{fluks} spectra as
dotted lines.  The left-hand panels of the figures show the synthetic spectra
constructed using the \cite{dbbr} STT relation, while the right-hand panels
show the analogous results when \cite{moz97}'s \teff\ scale is adopted.
The major uncertainty in evaluating the synthetic spectra through comparisons
such as those in Figures~\ref{tempcomp1} and~\ref{tempcomp2} is the reliability
of the MK spectral types of the ``intrinsic'' spectra of \cite{fluks}.
These authors obtained spectra of field M~giants and assigned each a
spectral type on the Case system using the criteria of \cite{nv64}.  By
scaling and averaging the spectra of all M~giants within one-half subtype of
an integral spectral type, \cite{fluks} derived what they called an
``intrinsic'' spectral sequence for M0--M10 giants on the Case system.  To
derive the analogous MK sequence, they assigned each Case ``intrinsic'' spectrum
an MK type and then interpolated (and extrapolated) to get ``intrinsic'' spectra
for M0--M10 giants on the MK system.  The transformation between Case spectral
types and MK types was derived from the relation tabulated by \cite{blanco}
between Mt. Wilson and Case spectral types, assuming that Mt. Wilson types and
MK types are identical (\cite{fitz69}; \cite{mikami}).  However, the latter 
assumption is questionable at early-M types (\cite{wing79}).  In addition, 
Blanco's transformation equates an M0~giant on the Case system to an M1.4~giant
on the MK system, so extrapolation of \cite{fluks}'s observational data was
required to produce their ``intrinsic'' MK spectra of
M0 and M1 giants.  For these reasons, we proceed cautiously when comparing
our synthetic spectra to \cite{fluks}'s spectral sequence, especially at
early-M spectral types, but we have nevertheless found their data
to be extremely useful in guiding us toward improving our synthetic spectra of
M~giants and determining the effective temperatures of these stars.  Unless
otherwise specified, all further references to \cite{fluks}'s ``intrinsic''
spectra can be assumed to mean those on the MK system.

As discussed in Section 2.2,
the continuum-forming regions of the stellar atmosphere are deep enough to be
unaffected by sphericity, and we therefore expect the synthetic spectrum
which has the same effective temperature as a star of a given spectral type
to match the ``continuum'' (i.e., inter-TiO) regions of the observed spectrum of
that star.  We also expect the main differences between the empirical spectra
and the synthetic spectra to be due to missing spectral lines in the synthetic
spectra.  This then implies that the synthetic spectrum of the correct
\teff\ should not have a lower flux than the corresponding ``intrinsic''
spectrum over any extended wavelength regime.

At spectral type~M0, where the \cite{moz97} and \cite{dbbr} temperature scales
only differ by 25~K, a 3880~K synthetic spectrum indeed proves to be a
reasonable fit to these ``continuum'' regions in the \cite{fluks} spectrum of
an average M0~giant.  In fact, for spectral
types M0--M3, the ``continuum'' region extending from 7300~to 7600~\AA\ in the
\cite{fluks} spectra is well-matched by the synthetic spectra calculated from
the \cite{dbbr} STT relation; for later types, this region is depressed
in the \cite{fluks} spectra, with respect to the respective \cite{dbbr}
synthetic spectra, probably due to the appearance of TiO and/or VO
absorption in the field stars.  Figure~\ref{tempcomp1} also suggests that the
\cite{dbbr} effective temperature for spectral type~M1 is perhaps a bit too hot,
but this could be due to errors in the aforementioned extrapolation of the
\cite{fluks} spectra as well.

Alternatively, while the synthetic spectrum expected to represent a given
spectral type on the \cite{moz97} scale often produces a good fit to the
inter-TiO regions blueward of 7000 \AA, the poorer agreement at redder
wavelengths makes these fits less satisfactory than the \cite{dbbr} results
for two reasons.  First, TiO absorption is expected to affect the flux
in the bluer regions of the spectrum for spectral types as early as~K5,
so there may not be any actual continuum points at visual wavelengths.
Second, even though the \cite{moz97} synthetic spectra fit the bluer
pseudo-continuum, the fact that they are cooler than the corresponding
\cite{dbbr} models means that they have a flux deficit over much of the
spectrum for $\lambda~>$~7000~\AA.  For these reasons, we conclude that
the \cite{moz97} temperature scale does not describe field M~giants.

Why doesn't the \cite{moz97} STT relation produce models which
agree with the field M~giant observations, when the agreement between the
\cite{moz97} and \cite{bg89} angular diameters for hotter stars is so
remarkable?  One could
conclude that something is wrong with the \cite{moz97} angular diameter
measurements, implying that the \cite{bg89} temperatures are also incorrect.
However, we propose instead that the temperature errors are not caused by faulty
\pud\ measurements but are due to incorrect estimates of the
limb-darkening correction.

Note that \cite{moz91} and Mozurkewich (1997) measure angular diameters at
8000~\AA, where the emergent flux of M~stars is affected by TiO absorption (see
Figures~\ref{tempcomp1} and~\ref{tempcomp2}).  However, TiO was not included in
the models of \cite{man79} which \cite{moz91} used to calculate limb-darkening
corrections.  Since TiO forms much higher in the stellar atmosphere than
the layers in which the continuum is produced, the radiation at 8000~\AA\ comes
from a part of the atmosphere much closer to the surface of the star than that
seen at continuum wavelengths.  This means that a smaller limb-darkening
correction is called for at 8000~\AA\ in M~giants than is appropriate at
continuum wavelengths.

Consequently, the limb-darkened angular diameters of the \cite{moz91} M~giants,
which we subsequently used to derive the corrections for the \cite{moz97} data,
are too large and result in effective
temperatures for these stars which are too cool (see equation~1).  However,
for hotter stars in which the TiO absorption at 8000~\AA\ is negligible,
the limb-darkening corrections are correct.  This explains
the good agreement between the \cite{moz97} and \cite{bg89} angular diameters,
since \cite{bg89} observed only G~and K~giants.  Assuming that the STT relation
of \cite{dbbr} given in Table~\ref{tefftable} is accurate for solar-metallicity
M~giants, we can force the \cite{moz97} temperature scale to match \cite{dbbr}'s
by altering the limb-darkening corrections which we applied to \cite{moz97}'s
data.  In this way, the limb-darkening corrections which should be applied to
uniform-disk angular diameters measured at 8000~\AA\ can be estimated.
For K~giants, the equation derived previously, \pld/\pud~=~1.078~+~0.002139~SP,
is appropriate; for M0--M4 giants, we suggest \pld/\pud~=~1.059~--~0.03317~SP;
and for giants later than type M5, a constant value, \pld/\pud~=~0.906, can be
used.

\section{The M Giant Models}

Based upon the comparison discussed in the previous section, we have chosen to
use the STT relation of \cite{dbbr} to model M~giants.  The second column of
Table~\ref{tefftable} gives the effective temperature and surface gravity (in
parentheses) which we assign to K~and
M~giants of a given spectral subtype.  The next step in constructing
representative M~giant synthetic spectra is to include TiO in the calculations,
adjusting the TiO band strengths as necessary
to try to match the spectra of field M~giants.

As mentioned previously, we have used the ``intrinsic'' MK spectra of M~giants
published by \cite{fluks} to evaluate our synthetic spectra.  In this process,
we have concentrated upon the models of M2--M5 giants.  We
discount the stars cooler than this because they are probably variable, and
their spectra are significantly affected by absorption bands of molecules not
included in our ODFs and/or spectral line lists, notably VO and H$_2$O.
At spectral types~M0 and~M1, the \cite{fluks} spectra
are more uncertain than those of later types because they are
extrapolations of the observational data.

\begin{figure}[p]
\epsfxsize=6.0in
\vspace*{-0.3in}
\hspace*{0.25in}
\epsfbox{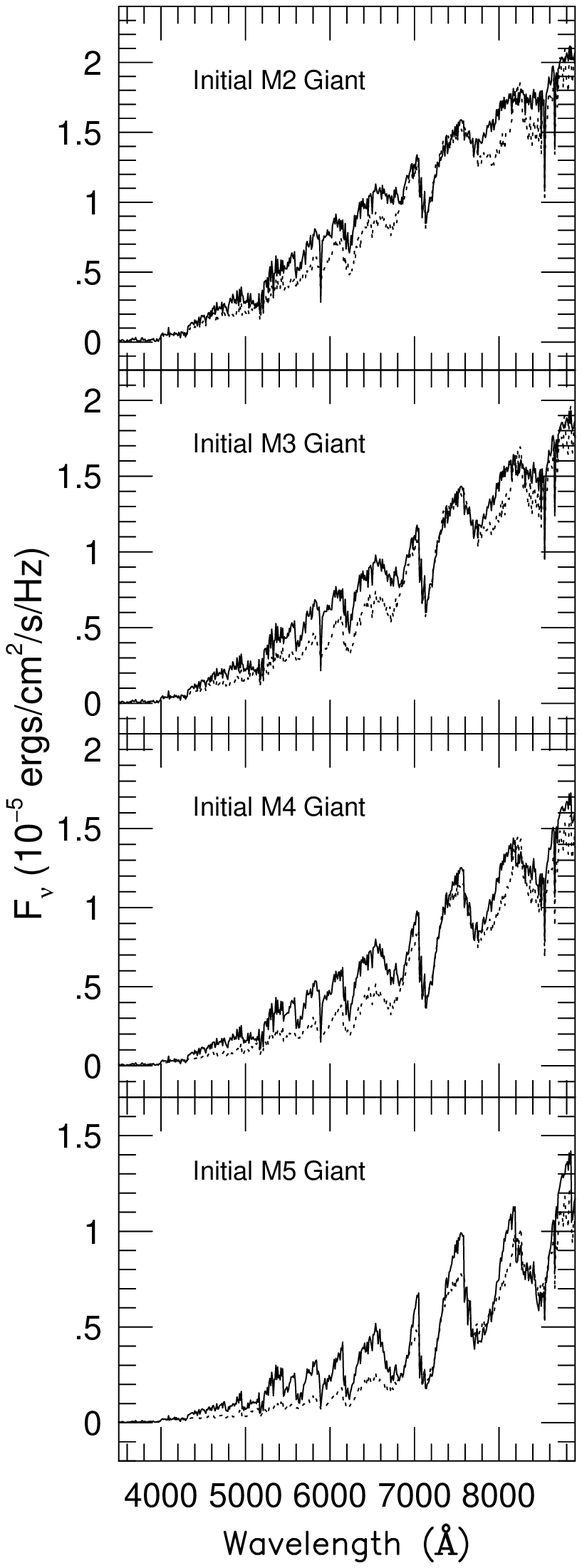}
\vspace*{-0.5in}
\caption{Comparisons of field-star spectra of M2--M5 giants
and our synthetic spectra which were calculated with our original molecular
data for TiO.  The solid lines are the solar-metallicity synthetic spectra,
and the dotted lines are the ``intrinsic'' MK spectra of field M~giants from
Fluks et al. (1994).}
\label{origspec}
\end{figure}

Our synthetic spectra of M2--M5 giants, calculated with the TiO line
list and original TiO molecular data described in Section 2.1, are
compared to the \cite{fluks} spectra in Figure~\ref{origspec}, where the
synthetic spectra are again represented by solid lines and the observational
data by dotted lines.  In this figure, we see that
the strength of the $\gamma$-system TiO band near 7100~\AA\ matches the
observed depth quite well, but the agreement for most of the other TiO bands
is much less satisfactory.
This has led
us to adjust the TiO data used in the calculations to attempt to produce
better agreement with the observational data.  These adjustments and the
resulting spectra are discussed in the following section.

\subsection{Treatment of TiO}

When performing spectral synthesis, especially for abundance analyses, it is
common to determine ``astrophysical'' oscillator strengths (i.e., gf~values)
for individual spectral lines.  For example,
if a line of an element of known abundance is stronger or weaker in the
synthetic solar spectrum than it is in the observed spectrum of the Sun, the
gf~value for that line is often adjusted
until the appropriate line strength is achieved.
We have chosen to use a similar approach to model the TiO bands in our
synthetic spectra of M~giants.

The band absorption oscillator strength for the 0--0 transition
of each system of TiO (f$_{00}$ in the notation of \cite{larsson}) is less
well-known than the Franck-Condon factors and the H\"{o}nl-London factors of
the various TiO lines.
Consequently, we have revised the
f$_{00}$ values of the TiO systems to reproduce the TiO band strengths seen in
the \cite{fluks} spectra at a given effective temperature.  For the redder
bands of TiO, which lie at least partially outside the wavelength regime
covered by the \cite{fluks} spectra, we have also used the spectra of
Terndrup et al.  (1990; hereafter \cite{tfw90}) and \cite{tfw91} as guides in
adjusting the TiO band strengths.
Because many of the absorption bands seen in the spectra of M~giants are made
up of overlapping systems of TiO, a set of synthetic spectra were first
calculated, each of which contained spectral lines from only one system
of TiO.  By isolating spectral features (or parts of features) which were
dominated by bands from a single system of TiO, we were able to unambiguously
adjust the f$_{00}$ values of the individual systems.
Simply for reference, Table~\ref{tiof00table}
compares the f$_{00}$ values which we eventually adopted to some others found
in the literature.  The resulting synthetic spectra of M~giants are shown in
Figure~\ref{finalspec}, where we compare our final spectra
to the \cite{fluks} spectra.  The agreement here is a significant improvement
over that seen in Figure~\ref{origspec}, especially for $\lambda$~$<$~7000~\AA.
However, some discrepancies remain and merit further discussion.

\begin{figure}[p]
\epsfxsize=5.5in
\vspace*{-0.7in}
\hspace*{0.5in}
\epsfbox{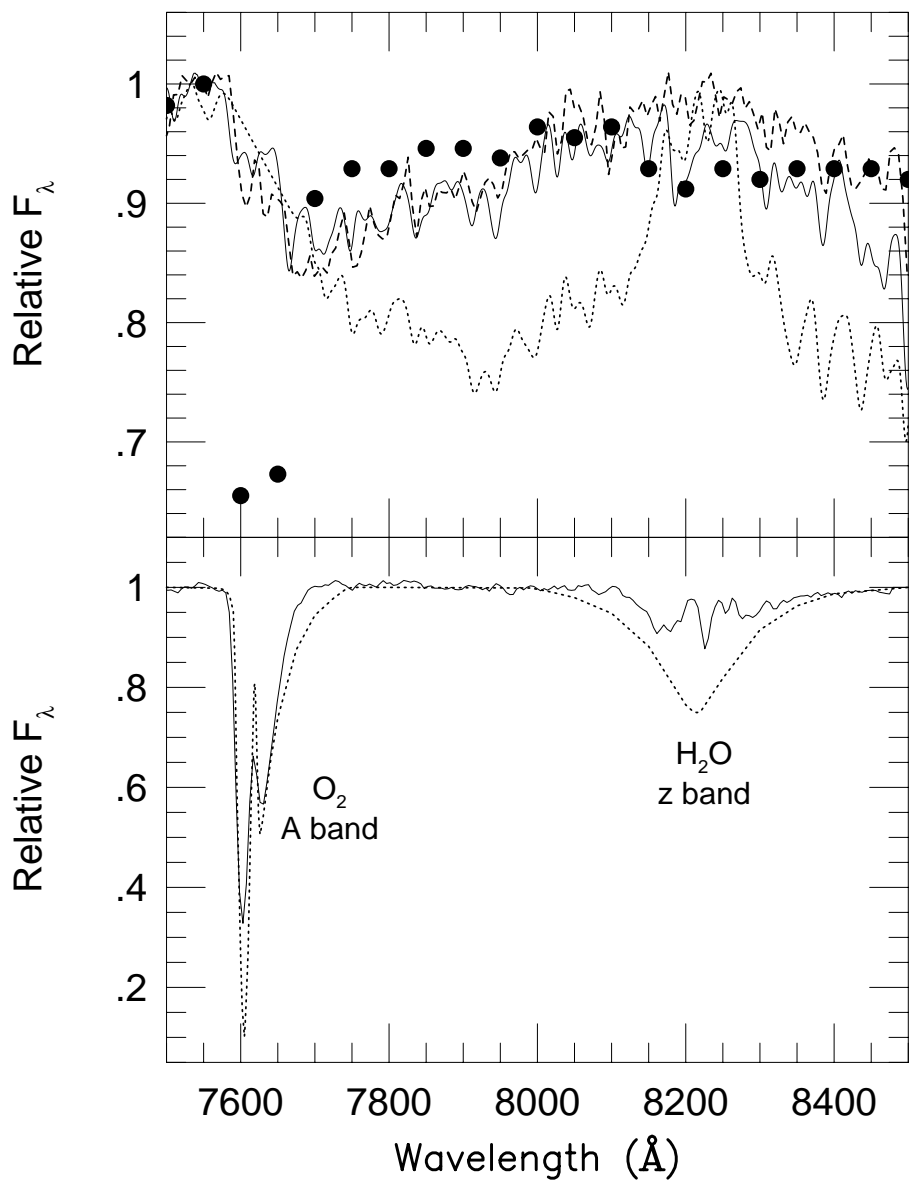}
\vspace*{-0.3in}
\caption{Illustration of the possible flux calibration error in
the Fluks et al. (1994) ``intrinsic'' spectra.  The upper panel compares the
M2~giant spectrum of Fluks et al. (dotted line) to the analogous synthetic
spectrum (solid line), the spectrum of HD~100783 (dashed line), a field
M2~giant observed by Terndrup et al. (1990), and the spectrum of HR~4517
(points), a field M1~giant observed by Kiehling (1987); the Fluks et al.
spectrum and the synthetic spectrum have been convolved to the resolution of
the Terndrup et al. data, and all of the spectra have been normalized near
7532~{\AA}.  In the lower panel, the telluric corrections
applied to the observational data by Fluks et al. (dotted line) and to the
HD~100783 spectrum by Houdashelt (1995; solid line) are shown; Kiehling's
spectrum has not been corrected for telluric absorption.}
\label{tellabs}
\end{figure}

There are large differences between the \cite{fluks} spectra and our synthetic
spectra in the wavelength region of 7600--8500~\AA, especially for the
earliest M~types.  We suspect that this discrepancy is due to an
error in the \cite{fluks} data, perhaps due to flux-calibration errors or
miscorrections for telluric absorption, since the synthetic spectra
calculated by \cite{fluks} showed similar systematic differences from their
``intrinsic'' field-star spectra.
In the upper panel of Figure~\ref{tellabs}, we support this
proposition by comparing our synthetic spectrum of an M2~giant (solid line) to
\cite{fluks}'s analogous spectrum (dotted line), the spectrum of the
M2~III star, HD~100783 (dashed line), observed by \cite{tfw90}, and the spectrum
of HR~4517 (points), a field M1~giant observed by \cite{kiehling}.  The lower panel shows the telluric corrections
applied to the observational data by \cite{fluks} (dotted line) and by
Houdashelt (1995) to the \cite{tfw90} spectrum (solid line); the {\it z}~band of
telluric H$_2$O centered near 8200~\AA\ clearly influences the region of
interest.  It is also clear from
this figure that the general shape of the TiO absorption in this
region of the synthetic spectrum is quite similar to that seen in HD~100783
and HR~4517.  At other wavelengths, the spectra
of \cite{fluks}, \cite{tfw90} and Kiehling are in much better accord.
Note also that the disagreement between the synthetic spectra and the
\cite{fluks} spectra near 8000~\AA\ in Figure~\ref{finalspec}
decreases for later spectral types, possibly because the TiO absorption begins
to dominate the telluric contamination.

Nevertheless, Figure~\ref{finalspec} shows that there are other
spectral regions, mostly located between the synthetic TiO bands, in which
the calculations are evidently missing some source of opacity, since
the synthetic spectra are brighter than the field star spectra there; these
occur near 5400, 6500, 7000 and 7500~\AA.
To explore the possibility that some of the missing absorption could be
supplied by higher-order rotational-vibrational lines of TiO than those
included in our line list, we computed synthetic
spectra using the TiO line lists of \cite{jorg94}; these include all lines
up to J=199 with $\upsilon^\prime$ and $\upsilon^{\prime\prime}$ values
between 0 and 10.  However, the use of J{\o}rgensen's line list did not produce
a noticeable difference in the
depths and morphology of the TiO bands in the optical region of the spectrum.
The $\alpha$ and $\beta$ systems of the two line lists were indistinguishable
in the synthetic spectra
and only minor differences were apparent for the $\gamma$ and $\gamma^\prime$
systems; the reddest bands of the latter two systems had bandheads which
were sharper and bluer but fit the observed spectra less well when
J{\o}rgensen's data were used.  For the $\delta$ system, on the other hand, the
TiO bands computed from J{\o}rgensen's line list had a more similar morphology
to those seen in the observational data, being very wedge-shaped, as opposed
to the more U-shaped bands produced by our line list.  While similar
differences were apparent in the shapes of the synthetic $\phi$-system bands,
we could not unambiguously detect these bands in any of
the empirical spectra, so no determination could be made regarding which line
list was more appropriate to use for this system.
Finally, the bandheads of J{\o}rgensen's $\epsilon$ system fall about
150~\AA\ bluer than predicted by the corresponding TiO line list of
\cite{plez96}, which produces a good match to the observational data.
Thus, the discrepancies between our synthetic spectra and the
\cite{fluks} field star spectra do not appear to be due to missing high-order
lines of the systems of TiO included in our spectral line list.

\cite{plez98} has kindly provided us
with plots of the a--f system of TiO and the spectrum of VO absorption in a
3300~K model.  It appears that most of the (inter-TiO region) differences
between our synthetic spectra and the field star spectra, as well as some of
those removed by altering the f$_{00}$ values of the other TiO systems, could 
be rectified by inclusion of these two absorption systems in the SSG spectral
line list.  However, the missing opacity near 6500~\AA\ does not appear
to be due to either TiO or VO, and until we can test these possibilities
further, we remain uncertain of the source of the missing opacity shortward of
7000~\AA\ in our synthetic spectra.  Keeping this caveat in mind, we
will proceed to examine our models further through more qualitative and
quantitative comparisons of spectral-type estimates, equivalent width
measurements and broad-band colors of the synthetic spectra and field M~giants.

\subsection{Molecular Bands in the Synthetic Spectra}

The main molecular species which influence the spectra of all M~giants are TiO
and CO.  Other molecules, such as CN, VO and H$_2$O, are also present in
the atmospheres of these stars, but their effects are important in a more
limited subset of the M~giants.  CN is most prevalent in the earliest M~types
but even then is often contaminated by overlapping spectral features due to
other molecules.  Absorption bands of VO and H$_2$O are seen only in spectral
types
M5~and later.  In the following, we discuss the CO and TiO band strengths
in our synthetic spectrum calculations and compare the results to observed
trends and to empirical spectra of field M~stars.

\subsubsection{TiO and Spectral Classification}

\begin{figure}[p]
\epsfxsize=6.0in
\vspace*{-0.3in}
\hspace*{0.25in}
\epsfbox{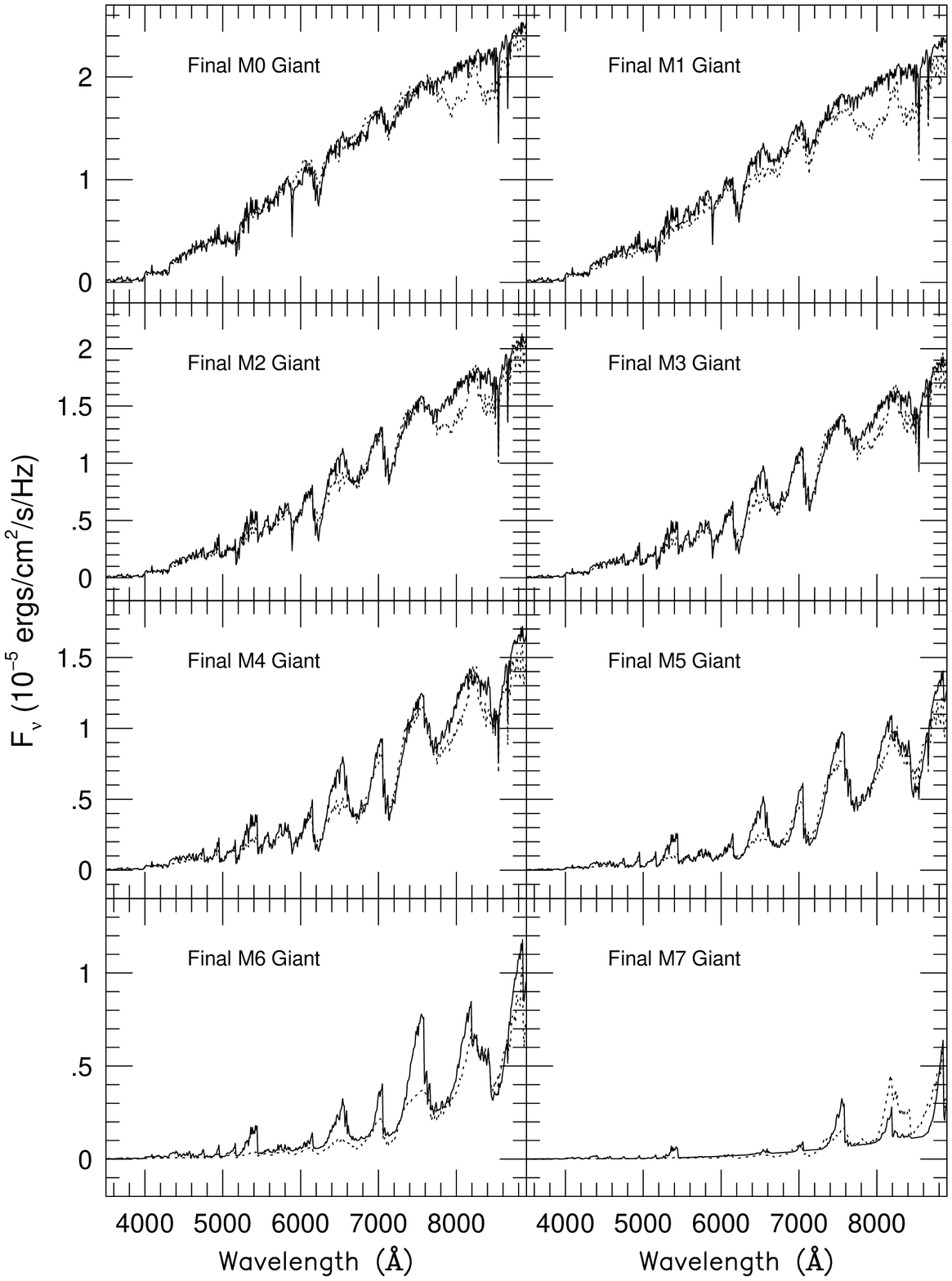}
\vspace*{-0.5in}
\caption{Comparisons of field-star spectra of M0--M7 giants
and our synthetic spectra which were calculated with our revised molecular
data for TiO.  The solid lines are the solar-metallicity synthetic spectra,
and the dotted lines are the ``intrinsic'' MK spectra of field M~giants from
Fluks et al. (1994).}
\label{finalspec}
\end{figure}

As mentioned previously, the spectral types of M~stars on the MK system are
determined from the strengths of the TiO bands.  For a given star, spectral
classification involves comparing the observed spectrum of the star to similar
spectra of standard stars which define the MK types, a somewhat qualitative
method for determining spectral classes which is not unlike the comparisons
we have made in Figure~\ref{finalspec}.  While this might lead us to conclude
that we have achieved our goal of reproducing the field~giant relation between
spectral type and effective temperature, it would be reassuring to be able to
verify this through something more robust than a fit-by-eye.  Thus, we have
also estimated spectral types from our synthetic spectra using three
quantitative measures of the TiO band strengths.

\begin{figure}[p]
\epsfxsize=6.0in
\vspace*{-0.4in}
\hspace*{0.25in}
\epsfbox{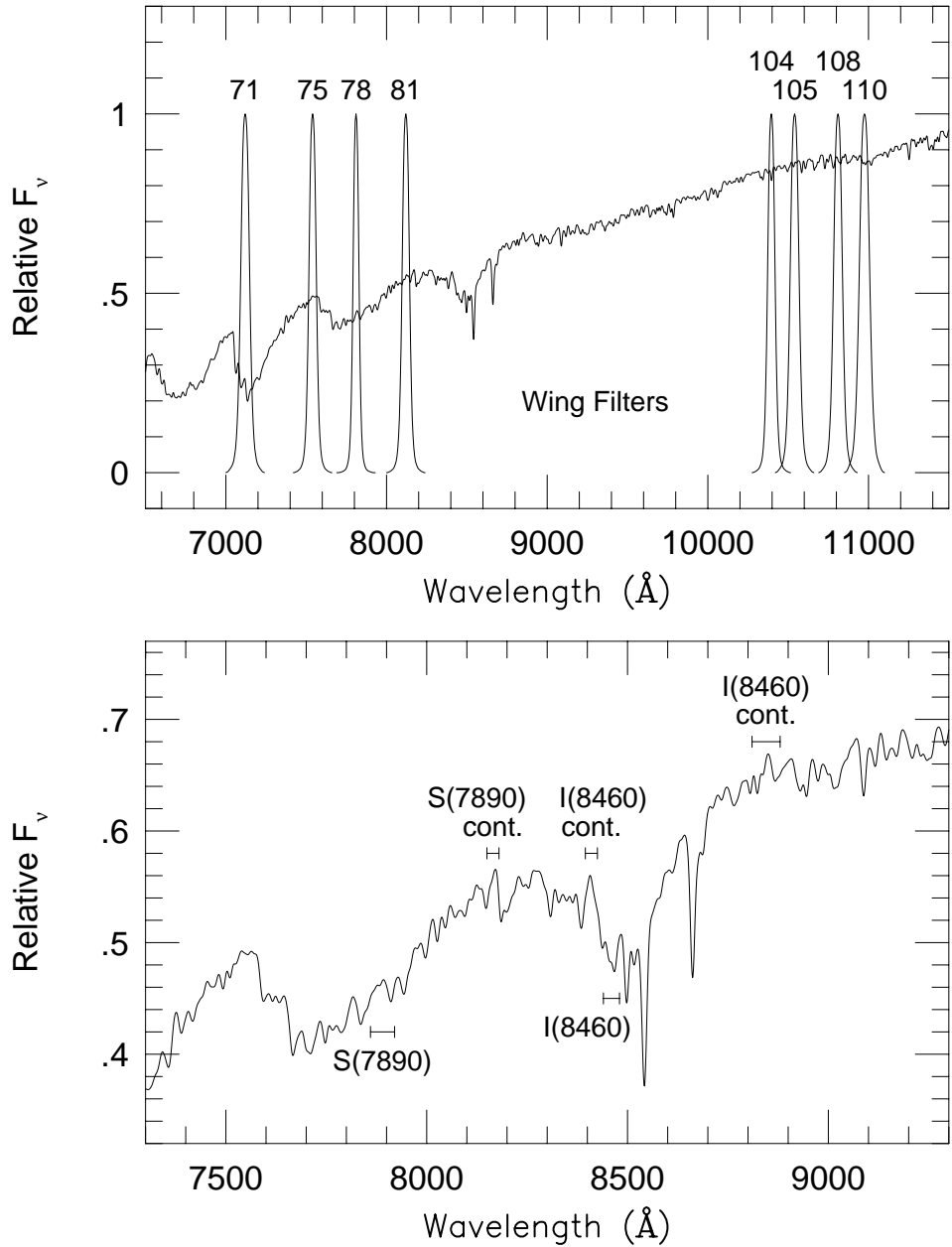}
\vspace*{-0.5in}
\caption{The systems used to estimate spectral types from TiO
bands in the synthetic spectra.  In the upper panel, the filter profiles of
Wing's (1971) photometric
system overlay our synthetic spectrum of an M3~giant.  In the lower panel, the
bandpasses used to define the S(7890) and I(8460) indices of Terndrup et al.
(1990), which measure pseudo-equivalent widths of TiO, are shown along with the
same synthetic spectrum which appears in the upper panel.}
\label{tiodefs}
\end{figure}

\cite{wing71} has designed an 8-color photometric system for determining
spectral types of late-K and M~giants; we illustrate the filter passbands
of this system
in the upper panel of Figure~\ref{tiodefs} along
with our synthetic spectrum of an M3~giant.  Wing's system uses the bluest of
his filters (filter~71 in Figure~\ref{tiodefs}) to measure the depth of the
band of the $\gamma$ system of TiO near 7100~\AA\ and estimate a spectral type
for the star, after correcting for the overlying CN absorption.
Since this method relies on a single TiO band, it is obviously applicable
only to stars for which this specific band is detectable and
is not saturated; this turns out to be spectral types~K4 through~M6.

We have measured synthetic colors on Wing's system and calculated spectral
types for our synthetic spectra using the methodology described by
\cite{mwc92}.  However, before determining these spectral types,
we had to calibrate the synthetic Wing colors to put them onto the
observational system.  This was done in a manner similar to that used to
calibrate the near-infrared, broad-band colors presented in \cite{hbs1}.
First, the zero point corrections
to be applied to the synthetic Wing magnitudes were determined from the
differences between the observed magnitudes of Vega (MacConnell et al. 1992)
and those measured from our synthetic spectrum of Vega (\cite{hbs1}).
After applying these zero-point corrections to the synthetic Wing magnitudes
of 35 of the field stars modelled in \cite{hbs1}, photometry of these stars
was used to derive linear relations between the photometric and
synthetic colors.  These calibration relations were then applied to the
synthetic Wing colors of the M~giant models before determining their spectral
types.

\cite{tfw90} defined a number of spectral indices (pseudo-equivalent widths)
which measure the strengths of TiO bands between 6000~and 8500~\AA.
For two of these indices, S(7890) and I(8460), Houdashelt (1995) presented
relationships between the index and spectral type for field stars of spectral
type~M1 and later.  The spectral regions used to define these two indices
are shown alongside our synthetic spectrum of an M3~giant in the bottom panel
of Figure~\ref{tiodefs}.  The S(7890) index measures the strength of
an absorption trough due primarily to a $\gamma$-system band of TiO
with respect to a single ``continuum'' sideband.  The I(8460) index measures
the strength of a bandhead of the $\epsilon$~system of TiO with respect to a
pseudo-continuum level interpolated from two adjacent spectral regions,
the bluer of which is affected by TiO absorption due to bands of both the
$\gamma$~and $\delta$~systems.  We have measured these indices from the
M-giant synthetic spectra shown in Figure~\ref{finalspec}
and used Houdashelt's relations to estimate their spectral types.

\begin{figure}[p]
\epsfxsize=6.0in
\vspace*{-0.7in}
\hspace*{0.25in}
\epsfbox{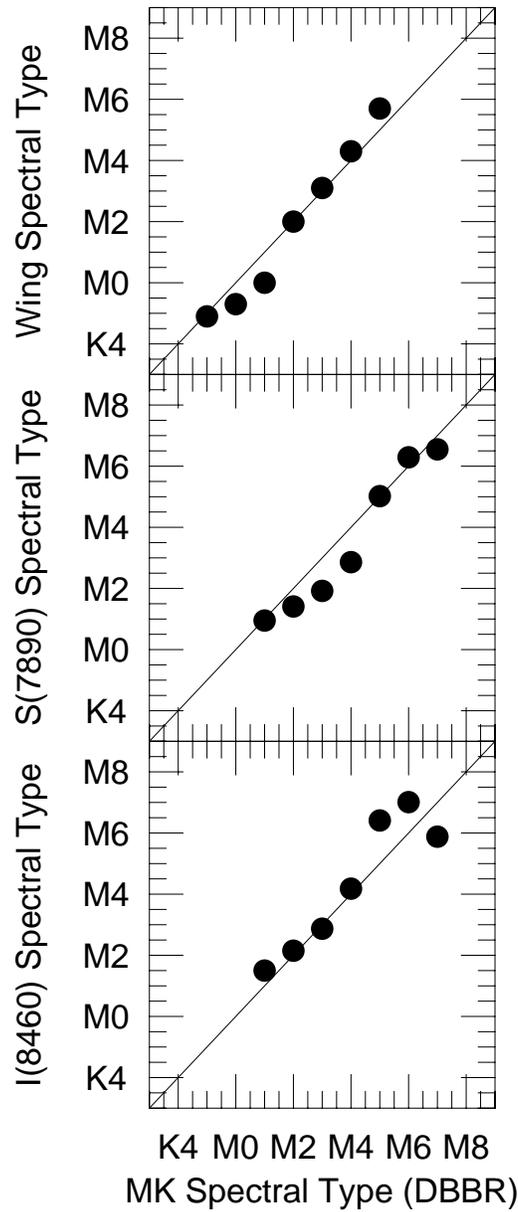}
\vspace*{-0.5in}
\caption{Comparison of the spectral types measured from
the TiO bands of the synthetic spectra and those indicated by their effective
temperatures (DBBR).  Solid lines show equality of the two spectral type estimates
in all panels.  The spectral types have been measured from the synthetic spectra
using Wing's (1971) photometric system (upper panel), adopting the
methodology of MacConnell et al. (1992); the S(7890) spectral index of
Terndrup et al.  (1990, center panel), using the the spectral type calibration
derived by Houdashelt (1995); and the I(8460) index of Terndrup et al.
(lower panel), using another of Houdashelt's calibrations.}
\label{tiotypes}
\end{figure}

In the upper panel of Figure~\ref{tiotypes}, we compare the spectral types
measured from our synthetic spectra using Wing's photometric system to those
implied by the effective
temperatures of the models per the \cite{dbbr} STT relation.  The agreement
is excellent, with the difference between the derived and expected spectral
types being larger than 0.3~subtypes only for spectral types~M0,~M1 and~M5.
Since the agreement is good at spectral type~K5, we suspect that the larger
discrepancies for the~M0 and~M1 stars occur because the effective
temperatures of the models are slightly too warm at these spectral types.
In effect, this simply means that \cite{dbbr}'s temperature estimate for an~M1
giant is a bit too hot.  Since they did not list \teff\ for spectral type~M0,
such a problem would affect the temperature at this spectral type as well
because we have taken a simple average of their~K5 and~M1 temperatures
to get \teff\ for type~M0.  Figure~\ref{tempcomp1}, which was consulted to
decide which effective temperature scale to adopt, supports the notion that
the \teff\ of a field M1~giant is hotter than that which \cite{dbbr} estimated
but implies that the M0~temperature cannot be too far off, again with the
caveat that the \cite{fluks} ``intrinsic'' spectra for spectral types~M0 and~M1
are extrapolations of their data and are therefore somewhat uncertain.
At spectral type M5, the difference between the Wing spectral type and that
inferred from the model's \teff\ may well be due to some of the
previously-discussed model uncertainties.  Judging from Figure~\ref{finalspec},
it is apparent that the ``continuum'' bands in our synthetic spectra,
especially the region near 7500~\AA\ which is integral in determining Wing
spectral types, are brighter than observed in the \cite{fluks}
spectra.  Since the 75 and 78 filters of Wing's system (see
Figure~\ref{tiodefs}) are definitely affected by VO absorption in late-M giants,
it is likely that
the Wing spectral types of the models cooler than about 3500~K are
later than those observed in the corresponding field giants mainly because
we don't include lines of VO in our spectral synthesis.

In the middle and lower panels of Figure~\ref{tiotypes}, we compare the
spectral types
derived from the S(7890) and I(8460) indices to the types assigned from the
effective temperatures of the models.  These diagrams can be broken into
two parts: early-M types (M1--M4) and late-M types (M5--M7).  For the former
group, the $\gamma$-system band of TiO with bandhead near 7600~\AA, which
dominates S(7890), is
probably a little too weak in our models (or there is insufficient flux in
the single continuum sideband), while the $\epsilon$-system bandhead
measured by the I(8460) index appears about right.  For the later types, the
S(7890) index gives spectral types agreeing with the effective temperatures,
but the I(8460) types appear to be too strong.  Unfortunately, since both the
S(7890) and I(8460)
indices lie in the region of the spectrum where we suspect \cite{fluks}'s
spectra to be in error (see Figure~\ref{tellabs}), the information
which can be gleaned about these indices through comparisons of the synthetic
spectra and the observational data is limited.  Nevertheless, we will briefly
discuss possible implications of these TiO index measurements in the synthetic
spectra.

It appears that the $\gamma$-system band of TiO which produces most of the
absorption measured by the S(7890) index may truly be too weak in our synthetic
spectra.  Because VO absorption would apparently have a greater affect on the
pseudo-continuum region used to measure S(7890) than on the spectral region of
the index itself (Plez 1998), this could be true even at later spectral types;
adding spectral lines of VO to the synthetic spectra would probably weaken the
S(7890) indices for models having \teff\ $\lesssim$~3500~K.  On the other hand,
making this band stronger by increasing the f$_{00}$ value of the
$\gamma$~system would throw off the good agreement between the synthetic spectra
and \cite{fluks}'s ``intrinsic'' spectra in other spectral regions; for example,
the absorption band with bandhead near 7100~\AA\ is also dominated by a
$\gamma$-system band, as is the redder half of the band between 6500~and
7000~\AA.  Thus, given the uncertainties in the calibration between S(7890) and
spectral type and the overall good agreement between the synthetic spectra
and the empirical data, we have chosen not to ``tweak'' this band to produce
S(7890) spectral types which agree better with those inferred by the model
temperatures.

Figure~\ref{finalspec} shows that the TiO absorption affecting the I(8460)
index does appear to grow more quickly with decreasing \teff\ in the synthetic
spectra than in the corresponding \cite{fluks} spectra.
However, the spectral region in which the I(8460) index is measured is highly
composite -- the sharp bandhead is due to a band of the $\epsilon$ system of
TiO, but there are overlying TiO bands from the~$\delta$ and~$\gamma$ systems
as well.  We again believe that adding spectral lines of VO to the synthetic
spectrum calculations would largely resolve the differences between the
I(8460) spectral types and the temperature-inferred types of the M5--M7 giants.

Even with the above considerations, the differences between the spectral types
estimated from the TiO bands of the synthetic spectra and those indicated by
their effective temperatures are small, always less than 1.2 spectral subtypes.
Thus, we conclude that the strengths of the TiO bands in our synthetic
spectra are in sufficient agreement with those expected from their effective
temperatures that we can be confident that the synthetic spectra which we
calculate will provide a good representation of M~giants in our evolutionary
synthesis models.

\subsubsection{CO and Spectral Classification}

One of the strongest absorption bands in cool giants is the first-overtone
$^{12}$CO(2,0)
band with bandhead near 2.3~$\mu$m.  In fact, \cite{bald} (1973) designed an
intermediate-band filter system specifically to measure the strength of this
band; these filters were later refined by \cite{cfp78} (1978) into the set
commonly used today.  \cite{bald} showed that
their CO index is a good luminosity indicator in cool stars, being stronger
in giants than in dwarfs of similar color; they found that the index
varies with \teff\ in giants as well, becoming stronger as \teff\ decreases.
\cite{bt91} have attributed the gravity behavior of this feature to
the lower continuous opacity in the atmospheres of cool giants with respect
to those of cool dwarfs of similar effective temperature; this effect more
than compensates for the lower abundance of molecules in the giant's atmosphere.

In a series of papers, the CO index was studied
in field dwarfs (\cite{paf77}), Galactic globular and open cluster members
(\cite{cfp78} 1978; \cite{fpc79}; \cite{pcmfa79}; \cite{cfpz80}; \cite{fpc81};
\cite{fpc83}), Magellanic Cloud clusters (\cite{pcmfa83}) and globular clusters
in M31 (\cite{fpc80}).  Frogel et al. (1978; hereafter \cite{fpam}) have
characterized the CO indices of field dwarfs and giants as a function of color,
and \cite{fbnmpa}, \cite{fpam},  and \cite{pcsmf} used the CO indices of
early-type galaxies and the nuclear region of M31 to infer that the (infrared)
integrated light of these objects must be giant-dominated.

Because this CO band is so gravity-sensitive, it is an
important stellar population diagnostic in integrated light studies, and CO
is a potentially informative spectral feature when applying our evolutionary
synthesis models to interpret observational data.  We expect to be able to
model the CO bands effectively in M~giants, since CO forms deep enough in the
stellar atmosphere to be relatively unaffected by extension and sphericity.
In fact, \cite{bb91} have shown that the behavior of the 2.3~$\mu$m CO band
with gravity and metallicity can be modelled quite reliably in G~and K~stars
with MARCS/SSG synthetic spectra, but we wish to
verify that this condition still holds for the current models, given the
improvements made to the ODFs and CO spectral line data (see Section 2.1 and
\cite{hbs1}) since their work.

\begin{figure}[p]
\epsfxsize=6.0in
\vspace*{-0.7in}
\hspace*{0.25in}
\epsfbox{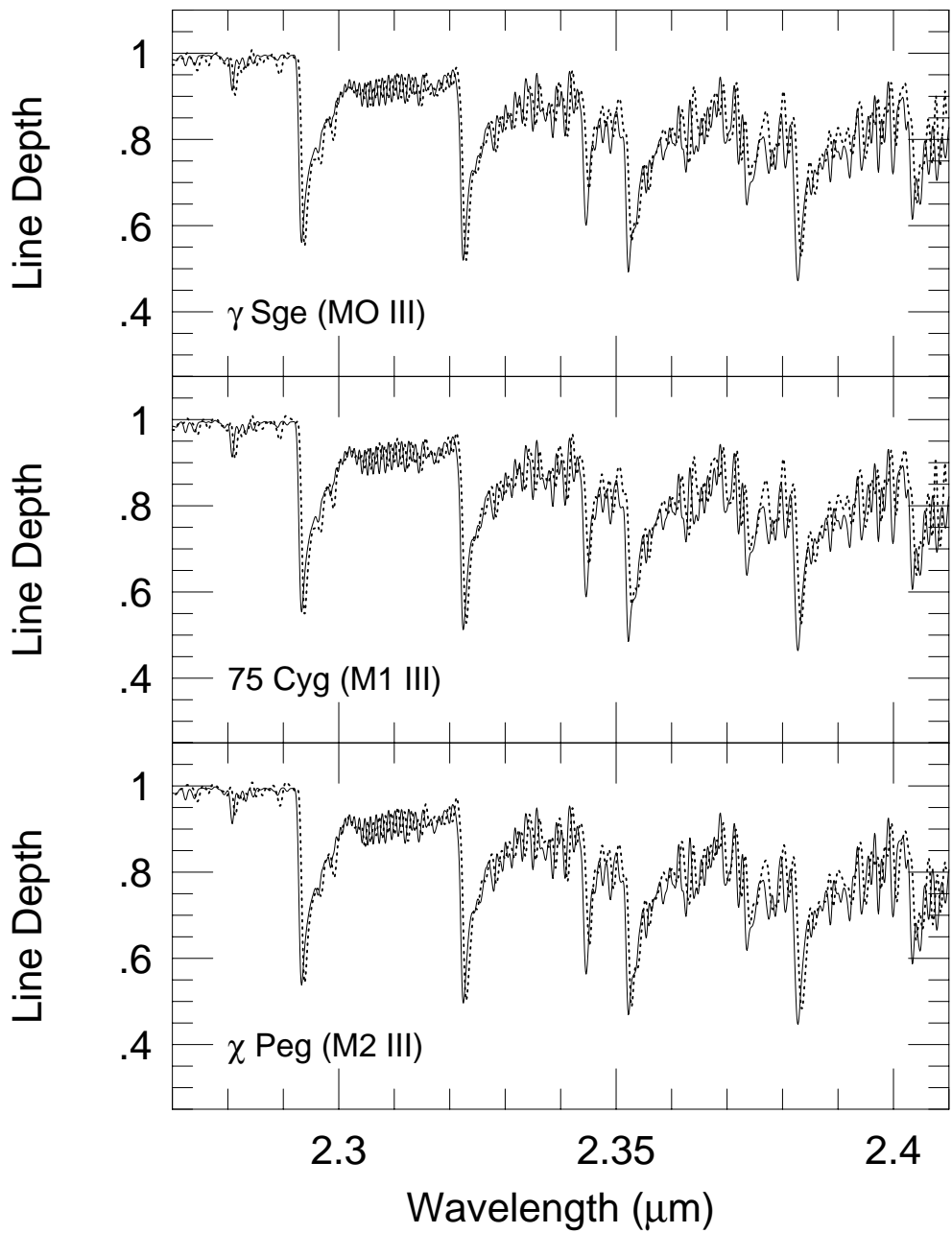}
\vspace*{-0.5in}
\caption{Comparisons of our synthetic spectra and the K-band
spectra of field M~giants observed by Kleinmann \& Hall (1986; KH86).
The bandhead of the $^{12}$CO(2,0) band is seen near 2.292~$\mu$m.
The solid lines represent the
synthetic spectra, and the dotted lines are the KH86 spectra.  The
synthetic spectra have been convolved to the resolution of the KH86
data and rebinned to 1.0~\AA\ pixels.  The slight wavelength difference
between the empirical and synthetic spectra occurs because the observed
spectra are calibrated using vacuum wavelengths, while the spectral
line lists used to calculate the synthetic spectra use wavelengths in air.}
\label{khcomp1}
\end{figure}

\begin{figure}[p]
\epsfxsize=6.0in
\vspace*{-0.3in}
\hspace*{0.25in}
\epsfbox{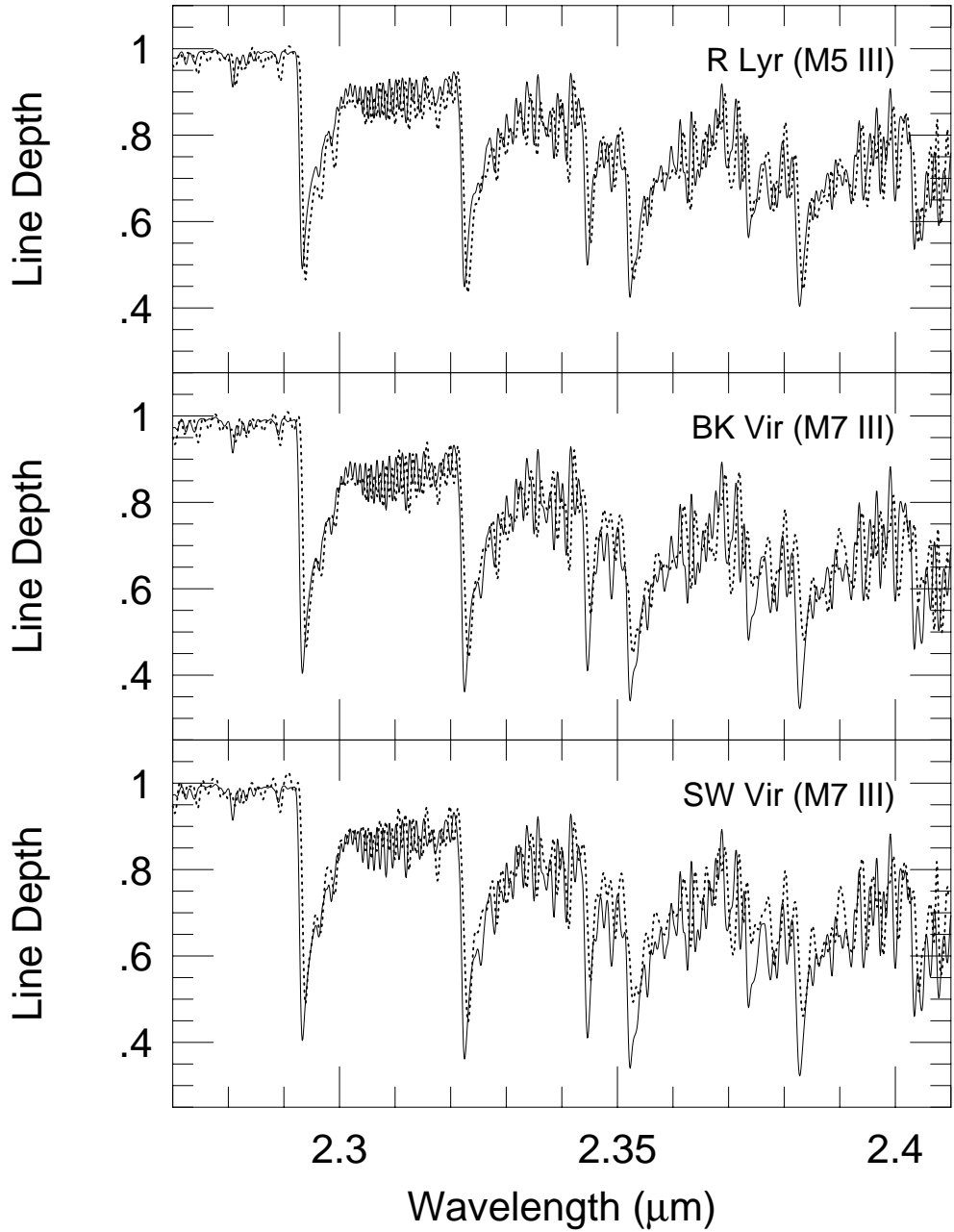}
\vspace*{-0.3in}
\caption{Further comparisons of our synthetic spectra and
the K-band spectra of field M~giants observed by Kleinmann \& Hall (1986).
The bandhead of the $^{12}$CO(2,0) band is seen near 2.292~$\mu$m.  See the
caption to Figure~10 for further details.}
\label{khcomp2}
\end{figure}

In Figures~\ref{khcomp1} and~\ref{khcomp2},
we compare our synthetic spectra of the appropriate spectral type to each of
the spectra of field M~giants observed by
Kleinmann \& Hall (1986; hereafter \cite{kh86}).  The agreement between the
synthetic spectra and the \cite{kh86} spectra is quite good throughout
the entire temperature range of M~giants, even at spectral type~M7, where the
field stars of \cite{kh86} are variable.  The $^{13}$CO bandheads at 2.345 and
2.365~$\mu$m can be seen in Figures~\ref{khcomp1} and~\ref{khcomp2} as well and
indicate that the $^{12}$CO/$^{13}$CO ratio varies among the
\cite{kh86} M~giants.

More recently, Ram\'{i}rez et al. (1997; hereafter \cite{ram97}) have developed
a scheme by which the
equivalent width of the 2.3~$\mu$m CO band, EW(CO), can be used to find
effective temperatures for K~and M~giants.  They obtained
spectra of field stars at low and intermediate resolutions (R=1380 and R=4830)
and measured EW(CO) for these stars, using slightly different continuum band
definitions at each resolution.  After converting spectral
type to effective temperature, \cite{ram97} found a linear relationship
between EW(CO) and \teff.  Because there did not appear to be any
systematic differences between their R=1380 and R=4830 measurements, they 
combined all of their data to derive this relation.

\begin{figure}[p]
\epsfxsize=5.0in
\vspace*{-0.8in}
\hspace*{0.75in}
\epsfbox{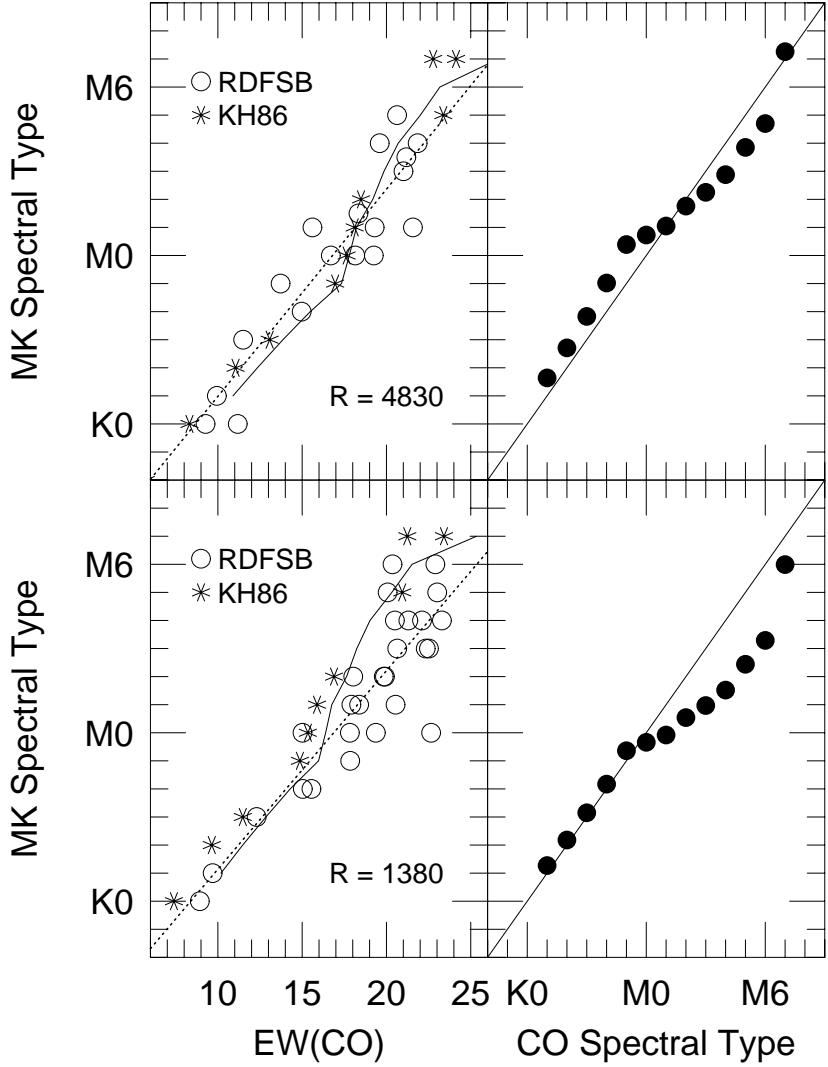}
\vspace*{-0.4in}
\caption{Comparison of the spectral types measured from
the CO bands of the synthetic spectra and those indicated by their effective
temperatures.  EW(CO) measures the depth of the CO bandhead near 2.3~$\mu$m
using passbands defined by Ram\'{i}rez et al. (1997; RDFSB); the continuum
passbands used are
resolution-dependent.  The upper panels show comparisons at a resolution, R, 
of 4830, the lower panels at R=1380.   In the left-hand panels, which show
the relationship between EW(CO)  and spectral type,
open circles represent the EW(CO) measurements of RDFSB, and
the dotted lines are linear, least-squares fits to this data.  The
asterisks show EW(CO) measured from the Kleinmann \& Hall (1986; KH86) spectra
using the appropriate continuum-band definitions,
and the solid lines connect the measurements of EW(CO) from our synthetic
spectra, after convolving and rebinning them to the resolution and dispersion
of the corresponding RDFSB spectra.  In the right-hand panels, the spectral
types estimated from the synthetic EW(CO) values (using the dotted
relations shown in the left-hand panels)
are compared to the spectral types indicated by the effective temperatures
of the models; the solid lines here represent equality of the two spectral type
estimates.}
\label{cotypes}
\end{figure}

We have convolved and rebinned our synthetic spectra of K~and M~giants to match
each resolution and dispersion of \cite{ram97} and have measured EW(CO) using
their continuum and CO band definitions.  We have also measured EW(CO) from the
\cite{kh86} spectra using both the R=4830 and R=1380 continuum definitions.
In Figure~\ref{cotypes}, we compare our measurements
of EW(CO) as a function of spectral type to those of \cite{ram97}.  In
this diagram, we have taken the spectral types of the \cite{ram97}
stars from SIMBAD, so they sometimes differ slightly from those adopted by
\cite{ram97}; for our synthetic spectra, we assume the spectral types
indicated from the STT relation of \cite{dbbr}.  Since our EW(CO) measurements
{\it do} show a resolution dependence, we plot the R=4830 results in the
upper panels of Figure~\ref{cotypes} and the R=1380 results in the lower panels.
The left-hand side of this diagram shows the relations between EW(CO) and
spectral type at each resolution; here, the data from \cite{ram97} is shown as
open circles, our \cite{kh86} measurements are asterisks, the solid lines
connect
the EW(CO) values measured from our synthetic spectra, and the dotted lines
are linear, least-squares fits to the \cite{ram97} data.  On the right-hand
side of the figure, the spectral types of the synthetic spectra, derived from
the dotted relations shown in the left-hand panels, are compared to the spectral
types inferred by \cite{dbbr}'s STT relation.

Some general conclusions can be drawn from Figure~\ref{cotypes}.  As expected
from the comparisons shown in Figures~\ref{khcomp1} and~\ref{khcomp2},
the EW(CO) measurements of the
synthetic spectra are in good agreement with those of the \cite{kh86} spectra,
with the CO absorption perhaps a bit stronger in the synthetic spectra than in the
\cite{kh86} spectra for the K~giants.
At R=4830, the synthetic spectrum CO widths and
\cite{ram97}'s CO widths agree well for the M~giants, with the CO again a bit
stronger in the K-giant synthetic spectra than in the observational data.
At R=1380, the situation is reversed from that observed at R=4830;
the synthetic spectra and the \cite{ram97} spectra produce
very similar EW(CO) for the K~giants, but in this case, the CO bands of
the synthetic spectra of the M~giants (and the \cite{kh86} spectra)
appear weaker than in the \cite{ram97} spectra.  Detailed comparisons
of the \cite{ram97} spectra and our synthetic spectra indicate that
these discrepancies are not due to differences in the strengths of the CO bands
but instead are caused by differences in the slope of the continuum just
blueward of the $^{12}$CO bandhead.

Overall then, given the scatter in the EW(CO) measurements of \cite{ram97},
and keeping in mind that a linear fit between EW(CO) and spectral type does not
appear to apply over the entire range of K~and M~spectral types, we conclude
that the spectral types estimated
for the synthetic spectra from their CO equivalent widths are in
agreement with those based upon their effective temperatures, especially at
R=4830.
Since K~and M~giants dominate the near-infrared light of most stellar
populations, Figures~\ref{khcomp1},~\ref{khcomp2} and~\ref{cotypes} also
show that our treatment
of the CO absorption in these cool stars will produce a realistic
representation of the CO bands in our evolutionary synthesis models.

\subsubsection{H$_2$O}

SSG has the option of including a spectral line list for water in the spectrum
synthesis calculations, but we have opted not to include H$_2$O absorption in
the models presented here.  This omission is based upon synthetic spectra
calculated using the water line lists of \cite{water1} and \cite{water2}.
Brett's line list is derived from the laboratory data of \cite{ludwig}
using the method described by \cite{plez92a}, while the list of \cite{water2}
is theoretical.

We first calculated models including spectral lines of H$_2$O using the line
list of \cite{water2}, but the water absorption seen in the
resulting synthetic spectra differed substantially from expectations based
upon observational data.  The H$_2$O lines depressed the flux much more evenly
throughout the infrared than is observed and did not show the familiar strong
bands which occur, for example, between the H and K atmospheric windows.  In
addition, the agreement between the observed and calculated CO bands was
substantially worsened in the coolest models which included these H$_2$O lines.
While the use of Brett's data produced spectra in which the water vapor
bands were more discrete, the bands which overlapped the CO absorption again
were sufficiently strong to spoil the nice agreement between the synthetic and
the empirical spectra seen in Figures~\ref{khcomp1} and~\ref{khcomp2}.

To date, we have not been able to determine whether the behavior of the H$_2$O
absorption in the models is due to problems with the stellar atmosphere models,
which possibly predict an overabundance of water, or the spectral line lists,
which may have oscillator strengths which are too large.  It is possible that,
to reproduce the spectrum of H$_2$O seen in real stars, the water absorption in
our synthetic spectra would require a treatment similar to that which we have
used for TiO.
Unfortunately, observational data from the Infrared Space Observatory, for
example, is not yet available to allow us to verify the need for such an
empirical calibration.  Nevertheless, since water absorption is not detectable
in the spectra of M~giants until spectral type~M5 or later (Bessell et al.
1989a),
the evolutionary synthesis models which we construct should not be significantly
affected by the omission of water lines in our synthetic spectra.

\subsection{Broad-Band Colors and Bolometric Corrections of K0--M7 Field Giants}

We have measured broad-band colors from our spectral-type sequence of K~and
M~giant synthetic spectra --
Johnson U--V and B--V; Cousins V--R and V--I; Johnson-Glass V--K, J--K and H--K;
and CIT/CTIO V--K, J--K, H--K and CO -- using the filter transmission profiles
described in \cite{hbs1}.  We have also computed CIT/CTIO K-band bolometric
corrections (BCs) for these models, assuming M$_{{\rm K},\odot}$~=~+3.31 and
BC$_{{\rm K},\odot}$~=~+1.41.  We present these colors and
BCs in Table~\ref{kmcolortable}; all of the colors have been
transformed to the observational systems using the color calibrations derived
in \cite{hbs1}.  Because these color calibrations are all very linear, we feel
comfortable extrapolating them into the regime of M~giant synthetic colors,
even though the sample
of field stars used to determine the calibration relations did not include
any stars cooler than spectral type~K5.  However, as in \cite{hbs1}, we
caution the reader that the U--V and H--K colors have greater uncertainties
than the other colors; the U--V colors are sensitive to missing opacity
in the ultraviolet region of the synthetic spectra, and the H--K color
calibrations are not well-determined (see \cite{hbs1}).

\begin{figure}[p]
\epsfxsize=5.5in
\vspace*{-0.8in}
\hspace*{0.5in}
\epsfbox{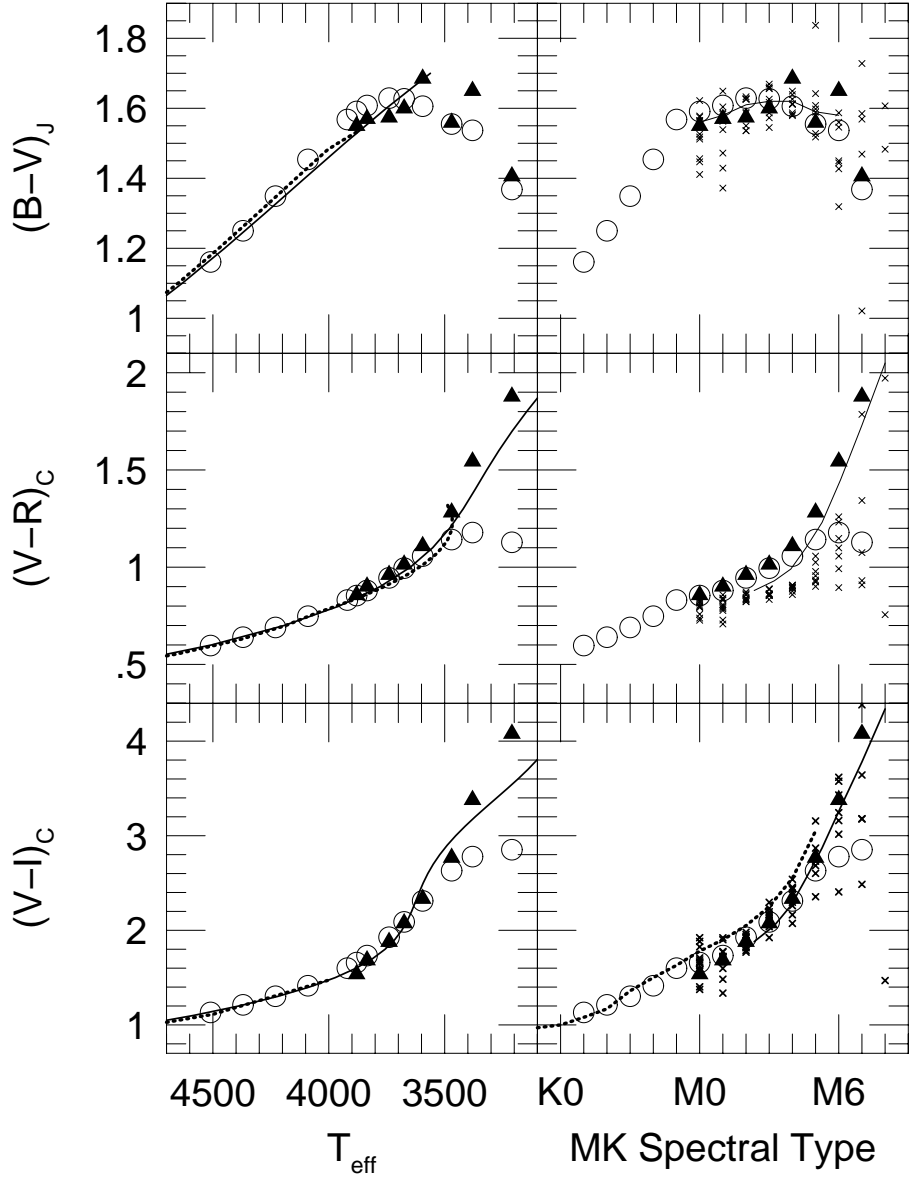}
\vspace*{-0.4in}
\caption{The calibrated, optical colors of the K~and M~giant models
are compared to field-giant color, \teff\ and color, spectral type relations.
The left-hand panels show the color, temperature comparisons; the right-hand
panels are color vs.~spectral type.  In all panels, the solid and dotted lines
are the field-star relations, the open circles represent our models, the
filled triangles show colors measured from the ``intrinsic'' MK spectra of
Fluks et al. (1994), and the small crosses are the photometry of field M~giants reported
by Fluks et al.  The field relations for color vs. \teff\ have been taken
from Bessell (1998; solid lines) and Gratton et al. (1996; dotted lines).
The color, spectral type field relations come from Lee (1970) for B--V,
Th\'{e} et al. (1990) for V--R and V--I (solid lines) and Bessell \& Brett (1988) for V--I (dotted line).  The
Case spectral types of Th\'{e} et al. have been converted to MK types using the
transformation given by Fluks et al.}
\label{opcolors}
\end{figure}

\begin{figure}[p]
\epsfxsize=6.0in
\vspace*{-1.5in}
\hspace*{0.25in}
\epsfbox{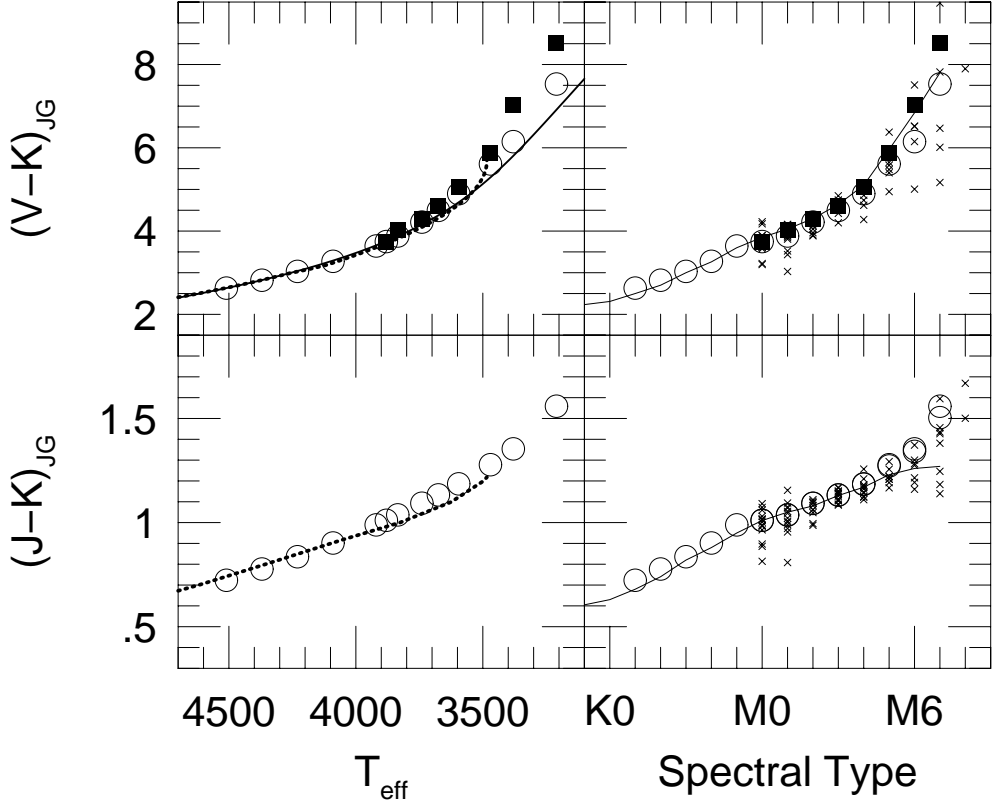}
\vspace*{-1.5in}
\caption{The calibrated V--K and J--K colors of the K~and M~giant models are
compared to field-giant color, \teff\ and color, spectral type relations.
The left-hand panels show the color, temperature comparisons; the right-hand
panels are color vs.~spectral type.  In all panels, the solid and dotted lines
are the field star relations, the open circles represent our models,
and the small crosses are the photometry of field M~giants reported
by Fluks et al. (1994).  Fluks et al.'s photometry has been transformed from the
ESO to the Johnson-Glass system using the color transformations given by
Bessell \& Brett (1988).  The field relations for color vs. \teff\ have been taken
from Bessell et al. (1998; solid line) and Gratton et al. (1996; dotted lines).  The color,
spectral type field relations come from Bessell \& Brett.  The filled squares
in the upper panels show the synthetic V--K colors which result when the V-band
magnitudes of the models are ``corrected'' for the missing opacity indicated
by the V~magnitudes measured from the corresponding Fluks et al. spectra.}
\label{ircolors}
\end{figure}

In Figures~\ref{opcolors} and~\ref{ircolors}, we compare our color vs.~\teff\ and color vs.~spectral type
relations to those observed for field M~giants.  We have extended these
comparisons into the K~giant regime to illustrate the general agreement
between the models and field relations for hotter stars.  In each of these
figures,
our models are represented by open circles, and the M~giant photometry
presented by \cite{fluks} is shown as small crosses.  We have also measured
colors directly from the ``intrinsic'' MK spectra of \cite{fluks} which we used
to calibrate the TiO bands, and these colors are shown as filled triangles
in Figure~\ref{opcolors}.  The field relations appear as solid and dotted lines;
their sources are described below and in the figure captions.
The filled squares seen in the upper panels of Figure~\ref{ircolors}
will be described as those specific panels are discussed.

Figure~\ref{opcolors} shows the optical color comparisons,
and the agreement between the models and the observational data is generally
quite good.  The color-temperature relations shown in the left-hand panels
of Figure~\ref{opcolors} have been taken from \cite{gcc96} (1996; dotted lines)
and Bessell (1998; solid lines); the latter were derived from the data of
\cite{bcp98}.  The field-star color, spectral type relations shown as solid
lines in the right-hand panels come from \cite{lee70} for B--V and from
\cite{the} for V--R and V--I, after first converting the latter's
Case spectral types to MK types using the transformation given by \cite{fluks};
the dotted relation in the V--I, spectral type panel is taken from Bessell \&
Brett (1988; hereafter \cite{bb88}).

The calibrated, optical, synthetic colors
and the colors measured from the \cite{fluks} spectra agree for the early-M
giants and then begin to diverge for the later types.  The B--V colors
of the \cite{fluks} spectra show a bit of random scatter, and the model B--V
colors may be a bit too red ($\sim$0.04 mag) for early-M giants, but the
synthetic colors fall well within the range of the \cite{fluks} photometry.
However, if we assume that any differences between the magnitudes measured
from the \cite{fluks} spectra and those measured from the models are due to
``errors'' in the model magnitudes, then a close inspection of the synthetic
magnitudes shows that the B--V colors of the models are about right only because
these ``errors'' in~B and~V largely offset one another.

In the V--R and V--I vs. \teff\ panels of Figure~\ref{opcolors}, the model
colors, the field relations and the colors measured from \cite{fluks}'s
spectra are essentially identical for spectral types earlier than
type~M4.  At cooler temperatures, the models and the \cite{fluks} spectrum
colors differentiate mainly because missing opacity in the synthetic spectra
makes their V-band magnitudes too bright.  For the coolest stars, it is
likely that variability, errors in the effective temperature determinations
and possibly small number statistics make the field giant color-temperature
relations and the \cite{fluks} ``intrinsic'' spectra less certain as well.

The analogous V--R and V--I vs.~spectral type comparisons are a little more
confusing.  First, the model colors and the \cite{fluks} spectrum colors again
agree to about spectral type M4 in both V--R and V--I, so we are doing a good job
of reproducing the \cite{fluks} spectra with our synthetic spectrum
calculations.  Second, \cite{fluks}'s R-band photometry is evidently not on
the Cousins system, since their V--R colors overlie neither the colors
measured from their spectra nor the model colors.  Finally, the field star
relations don't appear to be well-determined -- the \cite{the} and
\cite{bb88} field relations differ by $\sim$0.2 mag in V--I at a given spectral
type, and the two relations approximately bracket both the models and the
colors measured from the \cite{fluks} spectra.  This disagreement merits
some further discussion.

\cite{bb88} derived the field relation (dotted line) shown in the lower,
right-hand panel of Figure~\ref{opcolors} from a combination of V--I photometry
taken from \cite{cuz80} and spectral types taken from the Michigan Spectral
Survey (\cite{mss1}; \cite{mss2}; \cite{mss3}).  The relation of \cite{the},
the solid line, comes from their own photometry and Case spectral types derived
from their objective-prism spectra; we used \cite{fluks}'s transformation
from Case to MK spectral types to get the relation plotted, so it is perhaps
a bit more uncertain than \cite{bb88}'s relation.  Of course, if this
transformation is incorrect, then the MK spectral types of the ``intrinsic''
spectra of \cite{fluks} are also in error, and our treatment of the TiO bands
in the synthetic spectra is wrong as well.  Still, let us suppose that the
uncertainties in the transformation between Case and MK spectral types allow
a shift to earlier spectral types of the \cite{the} relation, the models and
the \cite{fluks} spectrum colors to make them agree with \cite{bb88}'s
field-giant relation.  Even then, \cite{fluks}'s photometry (crosses) would
not be similarly affected.  \cite{fluks} took the MK~spectral types of these
stars from the Bright Star Catalogue (\cite{bsc}), so to make these points lie
along \cite{bb88}'s relation requires systematic errors in \cite{fluks}'s
photometry; this is at least conceivable, given that their V--R colors appear
to be systematically too blue in the middle, right-hand panel of
Figure~\ref{opcolors}.

On the other hand, we also question whether \cite{bb88}'s field-giant relation
between spectral type and V--I is correct.  If we take the colors from this
relation and plug them into the V--I, \teff\ relation of \cite{bess98}, we get
an STT relation which differs significantly from that of \cite{dbbr}, which we
have used as the basis for our modelling of M~giants.  Thus, assuming that
\cite{bb88}'s V--I, spectral type relation is correct forces us to conclude
that either the color-temperature relation of \cite{bess98} is wrong or that
the STT relation of \cite{dbbr} is in error.  Without some additional
information, we cannot resolve these discrepancies between the V--I,
spectral type relations of field giants given by \cite{bb88} and \cite{the}.

Figure~\ref{ircolors} shows the comparisons for the V--K and J--K colors.
In this figure, the color, \teff\ relations of the field giants generally come
from the same sources as the optical relations; the V--K, \teff\ relation was
published by \cite{bcp98}.  The color, spectral type data for the field stars
is taken from \cite{bb88}.
The crosses again represent the photometry of \cite{fluks}; their ESO
colors have been transformed to the Johnson-Glass system using the color
transformations given by \cite{bb88}.  Since \cite{fluks}'s spectra only
extended to 9000~\AA, near-infrared colors could not be measured from them.

The color most often used to determine effective temperatures of cool stars is
V--K, so
it would be gratifying if our models predicted the same V--K, \teff\ relation
as that observed in field M~giants.  This appears to hold true for
spectral types earlier than about type~M4, but the cooler models become
progressively redder than the color-temperature relation of \cite{bcp98},
reaching $\sim$0.6 mag redder at spectral type M7.  However, the opposite holds
true for the V--K, spectral type data.  Here, the models are slightly bluer than
the field relation (\cite{bb88}) but overlap \cite{fluks}'s transformed
photometry.
The filled squares shown in the upper two panels of Figure~\ref{ircolors} show
the V--K colors of the models which result when the synthetic V--band magnitudes
are ``corrected'' for their differences with the respective V~magnitudes
measured from the \cite{fluks} spectra; this approximates the V--K colors we
would expect to measure from the \cite{fluks} spectra if their wavelength
coverage included the K~band.  This adjustment makes the model colors an
excellent
match to the field stars in the color, spectral type plane but obviously
makes the fit to the M-giant color-temperature relation much worse.  However,
we expect the synthetic V--K colors to be too red for spectral types~M5 and
later because we
have neglected H$_2$O absorption in our calculations; it's inclusion would
make the K magnitudes fainter but leave the V magnitudes unaffected.  Since
we encounter the same uncertainty here that we experienced in the V--I plots,
namely that substituting the colors from \cite{bb88}'s V--K, spectral type
relation into \cite{bcp98}'s V--K, \teff\ relation gives an STT relation which
differs from that of \cite{dbbr}, it is not clear to us which field relation
(color-temperature or color-type) is more reliable.  In either case,
the model vs. field-star color differences are relatively small for the
early-to-mid-M giants, so
we are satisfied that our models provide an adequate representation of
the V--K colors of field M~giants to be used for evolutionary synthesis.

The J--K colors of the models match the field relations for the K~giants but
become slightly redder than the field relation of \cite{gcc96} (1996) at a given
\teff\ for early-M~types.  The models, however, are in excellent agreement with
the J--K vs.~spectral type relation of \cite{bb88} through spectral type~M4.
As mentioned in Section 3.1, we have not been able to calibrate the absorption
bands of the $\phi$-system of TiO, some of which fall in the J~band, using
empirical spectra.  Therefore, we allowed the J--K colors of the models to
assist us in choosing a final f$_{00}$ value for the $\phi$~system, since the
model J--K colors redden significantly as this parameter is increased.
For the later M~types, we expect that adding spectral lines of H$_2$O to the
synthetic spectrum calculations, while diminishing both the J-band and K-band
fluxes, would resolve the remaining differences between the models and the
field-giant relations.

\begin{figure}[p]
\epsfxsize=5.0in
\vspace*{-0.8in}
\hspace*{0.75in}
\epsfbox{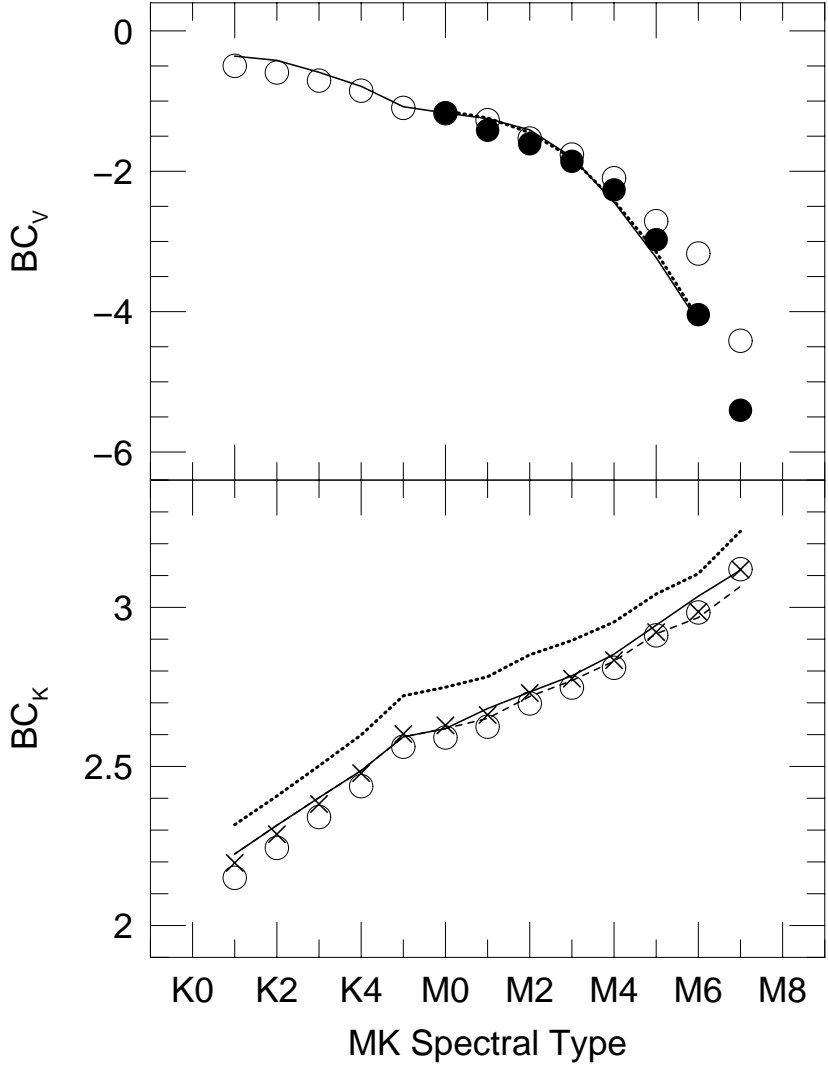}
\vspace*{-0.4in}
\caption{The bolometric corrections (BCs) of the K~and M~giant
models are compared to those of field giants.  The model BCs are shown as open
circles.  In the upper panel, the field-giant BC$_{\rm V}$, spectral type
relation of
Johnson (1966) is shown as a solid line; that of Lee (1970) is a dotted line.
The filled circles show the BCs of the models if the synthetic V-band magnitudes
are ``corrected'' for their differences with the V~magnitudes measured from the
``intrinsic'' MK spectra of field M~giants of Fluks et al. (1994).  The bottom
panel compares the model CIT/CTIO BC$_{\rm K}$ values to those predicted for field
giants of the same color using the relations of Bessell \& Wood (1984).  The
solid and dashed lines are derived from the synthetic (V--K)$_{\rm CIT}$ colors;
the latter results after ``correcting'' the V~magnitudes of the models for their
differences with the empirical data.  The dotted line and crosses are derived
from the synthetic (J--K)$_{\rm CIT}$ colors, after transforming them to the
AAO system using the color transformation of Bessell \& Brett (1988); the
former comes from Bessell \& Wood's solar-metallicity relation, while the
latter result when their relation for metal-poor stars is used instead.}
\label{bcplot}
\end{figure}

In Figure~\ref{bcplot}, we compare the bolometric corrections of our models of
field K~and M~giants, as a function of spectral type, to empirical relations.
Recall that the BCs of the models assume M$_{{\rm V},\odot}$~=~+4.84,
BC$_{{\rm V},\odot}$~=~--0.12, M$_{{\rm K},\odot}$~=~+3.31 and
BC$_{{\rm K},\odot}$~=~+1.41; this implies a (V--K)$_{\rm CIT}$ color for the
Sun of 1.53, which is $\sim$0.02 mag redder than the best observed values
tabulated by \cite{bcp98}.
In the upper panel of this figure, the open circles are the (untabulated) V-band
BCs of our models, the solid line is the field relation of \cite{johnson}, and
the M-giant relation of \cite{lee70} is shown as a dotted line.  The two field
relations are virtually identical and are in close agreement with the model
BCs for spectral types K0--M3; at later types, the models predict BC$_{\rm V}$
values which are smaller in magnitude than the field relations imply.  However,
recall that our synthetic V-band magnitudes are probably too bright, due to
missing opacity in the synthetic spectrum calculations.  If we substitute
the V-band magnitudes measured from the ``intrinsic'' MK M-giant spectra of
\cite{fluks} into the BC$_{\rm V}$ calculations, then the model points move to the
positions of the filled circles in
the upper panel of Figure~\ref{bcplot}; the latter are a much better match to
the field relations at late-M spectral types, showing that the differences
between the field-star BCs and the model BCs are consistent with the
corresponding differences in their observed and computed V~magnitudes.
It is precisely these uncertainties in the V-band magnitudes that led us to
tabulate BC$_{\rm K}$ in Table~\ref{kmcolortable} rather than BC$_{\rm V}$.

In the lower panel of Figure~\ref{bcplot}, we compare our K-band bolometric
corrections to some near-infrared field relations.  Here, the open circles
again represent our models, and the empirical trends have been calculated
as prescribed by \cite{bw84} (1984).  \cite{bw84}
give relations between BC$_{\rm K}$ (on the CIT/CTIO system) and both
(V--K)$_{\rm CIT}$ and (J--K)$_{\rm AAO}$.  Using the color transformation
between (J--K)$_{\rm CIT}$ and (J--K)$_{\rm AAO}$ from \cite{bb88}, we have used
the calibrated, synthetic V--K and J--K colors of our models to calculate the
BC$_{\rm K}$ values which the Bessell \& Wood relations predict; the results are
shown in the lower panel of Figure~\ref{bcplot}.  The solid line comes from
the model (V--K)$_{\rm CIT}$ colors, while the dashed line is produced when
these colors are
``corrected'' for the differences between the V-band magnitudes of the models
and those measured from the ``intrinsic'' M-giant spectra of \cite{fluks}.
The dotted line results from the synthetic J--K colors when Bessell \& Wood's
solar-metallicity J--K relation is used, while the crosses are the analogous
points derived from their metal-poor relation.  Surprisingly, the K-band BCs
of the models better match the predicted BCs of metal-poor field stars of
similar J--K color than those of their solar-metallicity counterparts.  However,
given the uncertainties in the calibration of the field-star BC$_{\rm K}$ vs.~J--K
relations and the nice agreement between the model BCs and those of field giants
of similar V--K color, we can confidently recommend the use of our 
color-temperature relations and BC$_{\rm K}$ values of K~and M~giants for converting
isochrones from log~\teff, log~L space into the color-magnitude plane.

\subsection{Color-Temperature Relations of M~Giants}

Given the generally good match between the broad-band colors and bolometric
corrections measured from
our synthetic spectra of field K~and M~giants and the empirical data,
we have proceeded to construct grids of models of cool giants to supplement
those presented in \cite{hbs1}.  At each of four metallicities, we have
calculated MARCS model atmospheres and SSG synthetic spectra for stars having
3000~K~$\leq$~\teff~$\leq$~4000~K and --0.5~$\leq$~log~g~$\leq$~1.5.
Table~\ref{gridtable} gives the calibrated colors and CIT/CTIO K-band
bolometric corrections of these models; column~3 of
Table~\ref{gridtable} gives the spectral types
measured from the synthetic spectra using the photometric system of
\cite{wing71}.  For those who are concerned about possible errors in the
synthetic spectra due to missing opacity and/or spectral lines, we also provide
Table~\ref{flukscomptable}, which gives the differences between the calibrated,
synthetic V~magnitudes and optical colors of our models of field M0--M7 giants
and those measured from \cite{fluks}'s ``intrinsic'' M-giant spectra.
The spectral types of the models can be used in conjunction with this table to
``correct'' the synthetic colors to match the observational data as desired,
but we urge the reader to thoroughly review \cite{fluks} before adopting these
color corrections.  Also, keep in mind that the color calibrations have been
derived from Population~I stars, so the colors of the models having
[Fe/H]~$\lesssim$~--0.5 should be used with some degree of caution.

\section{Conclusions}

To better model elliptical galaxies through evolutionary synthesis, we have
improved our synthetic spectra of M~giants by 1) determining the optimal
effective temperature scale to use for these cool stars, 2) adjusting the
f$_{00}$ values of the TiO bands to best match the band strengths observed in
the spectra of field M~giants, and
3) evaluating the resulting models by comparing the synthetic spectra,
their estimated spectral types and the model colors and bolometric corrections
to empirical data.

We have critically examined three effective temperature scales for M~giants,
each derived from angular diameter measurements.  Two of these were taken from
Dyck et al.  (1996; \cite{dbbr}) and Di Benedetto \& Rabbia (1987); the third
was derived from the angular diameters measured by \cite{moz91} and Mozurkewich
(1997; \cite{moz97} collectively).  We found that the effective temperature vs.~spectral type
relation of Dyck et al. (1996) produces synthetic spectra which have the
same continuous flux level as the ``intrinsic'' M~giant spectrum of the same
spectral type observed by Fluks et al. (1994).  A possible exception to this
rule occurs at spectral type~M1, where the Dyck et al. \teff\ may be a bit
too cool.  This temperature scale, which is similar to Di Benedetto \& Rabbia's
but covers a wider range of spectral types, also proves to be a good match to
that of Bell \& Gustafsson (1989; \cite{bg89}), which we adopted in a companion
paper discussing color-temperature relations of hotter stars (Houdashelt et al.
2000).  While the angular diameters measured by \cite{moz97} were found to
match those predicted by \cite{bg89} for G~and K~giants remarkably well, the
resulting effective temperature scale could not be reliably extended into the
M~giant regime because his uniform-disk angular diameters were measured at
8000~\AA.  At this wavelength, TiO absorption is present in M~star atmospheres,
and the limb-darkening corrections used by \cite{moz91} did not take this into
account.

Adopting \cite{dbbr}'s effective temperature scale, we have constructed MARCS
model atmospheres and SSG synthetic spectra for solar-metallicity K0--M7 giants.
For each system of TiO, we adjusted the band absorption oscillator strength
for the 0--0 transition, f$_{00}$, until we were best able to reproduce the
``intrinsic'' MK spectra of field M~giants of Fluks et al. (1994).  We found the
resulting synthetic spectra to be a good match to the K-band spectra of
Kleinmann \& Hall (1986) as well.  Quantitative measures of the spectral types
of the M~giant synthetic spectra based upon the strengths of both the
TiO bands and the CO bandhead near 2.3~$\mu$m are in good agreement with
the spectral types expected from \cite{dbbr}'s temperature scale.  In addition,
the broad-band colors of the K~and M~giant sequence are quite similar to
those expected of solar-metallicity field stars of the same spectral type and/or
\teff, especially for the K~and early-M stars.  At later spectral types, most
of the differences between the models and the empirical data can be ascribed
to our omission of spectral lines of VO and H$_2$O in the spectral synthesis.

Finally, we have presented colors and bolometric corrections for models having
3000~K~$\leq$~\teff~$\leq$~4000~K and --0.5~$\leq$~log~g~$\leq$~1.5 at four
metallicities: [Fe/H]=+0.25, 0.0, --0.5 and --1.0.  These supplement and
extend the color-temperature relations presented in our companion paper
(Houdashelt et al. 2000).

\acknowledgments

We would like to thank the National Science Foundation (Grant AST93-14931)
and NASA (Grant NAG53028) for their support of this research.  We also
thank Ben Dorman for allowing us to use his isochrone-construction code
and Mike Bessell for providing many helpful suggestions on the manuscript.
MLH would like to express his gratitude to Rosie Wyse for providing support
while this work was completed.
The research has made use of the Simbad database, operated at CDS, Strasbourg,
France.

\begin{table}
\dummytable\label{angdiamtable}
\end{table}

\begin{table}
\dummytable\label{tefftable}
\end{table}

\begin{table}
\dummytable\label{tiof00table}
\end{table}

\begin{table}
\dummytable\label{kmcolortable}
\end{table}

\begin{table}
\dummytable\label{gridtable}
\end{table}

\begin{table}
\dummytable\label{flukscomptable}
\end{table}

\clearpage

\begin{figure}[p]
\vspace*{-1.3in}
\hspace*{-1.0in}
\epsfbox{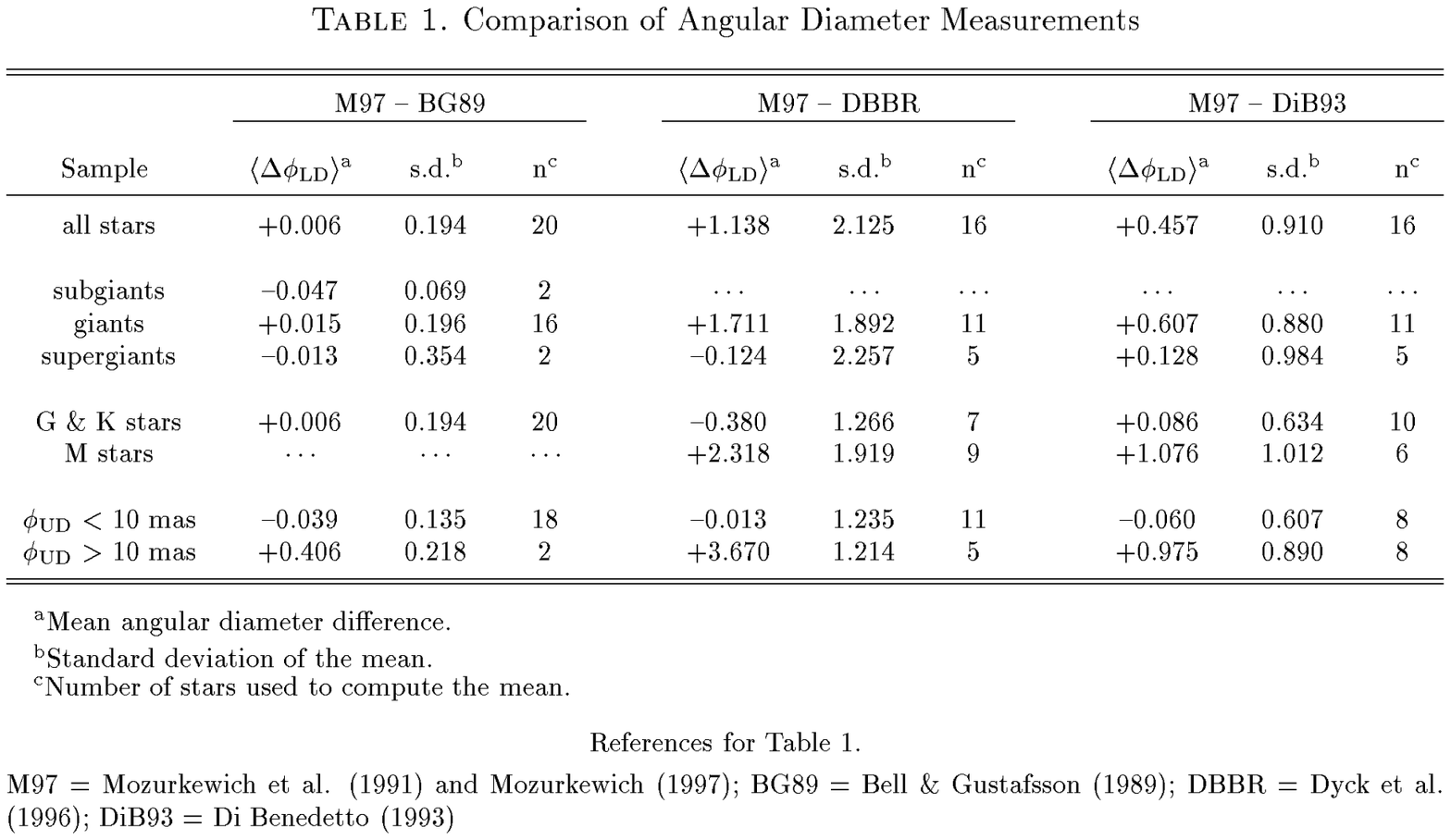}
\end{figure}

\begin{figure}[p]
\vspace*{-1.3in}
\hspace*{-1.0in}
\epsfbox{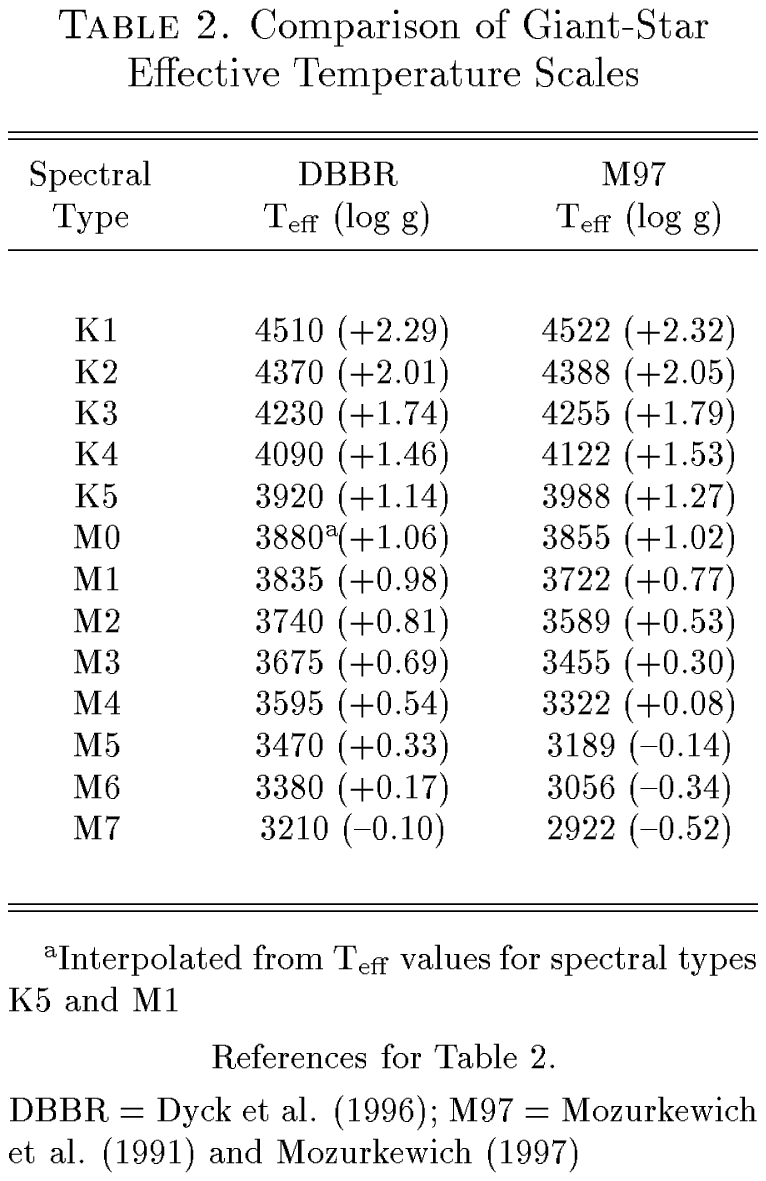}
\end{figure}

\begin{figure}[p]
\vspace*{-1.3in}
\hspace*{-1.0in}
\epsfbox{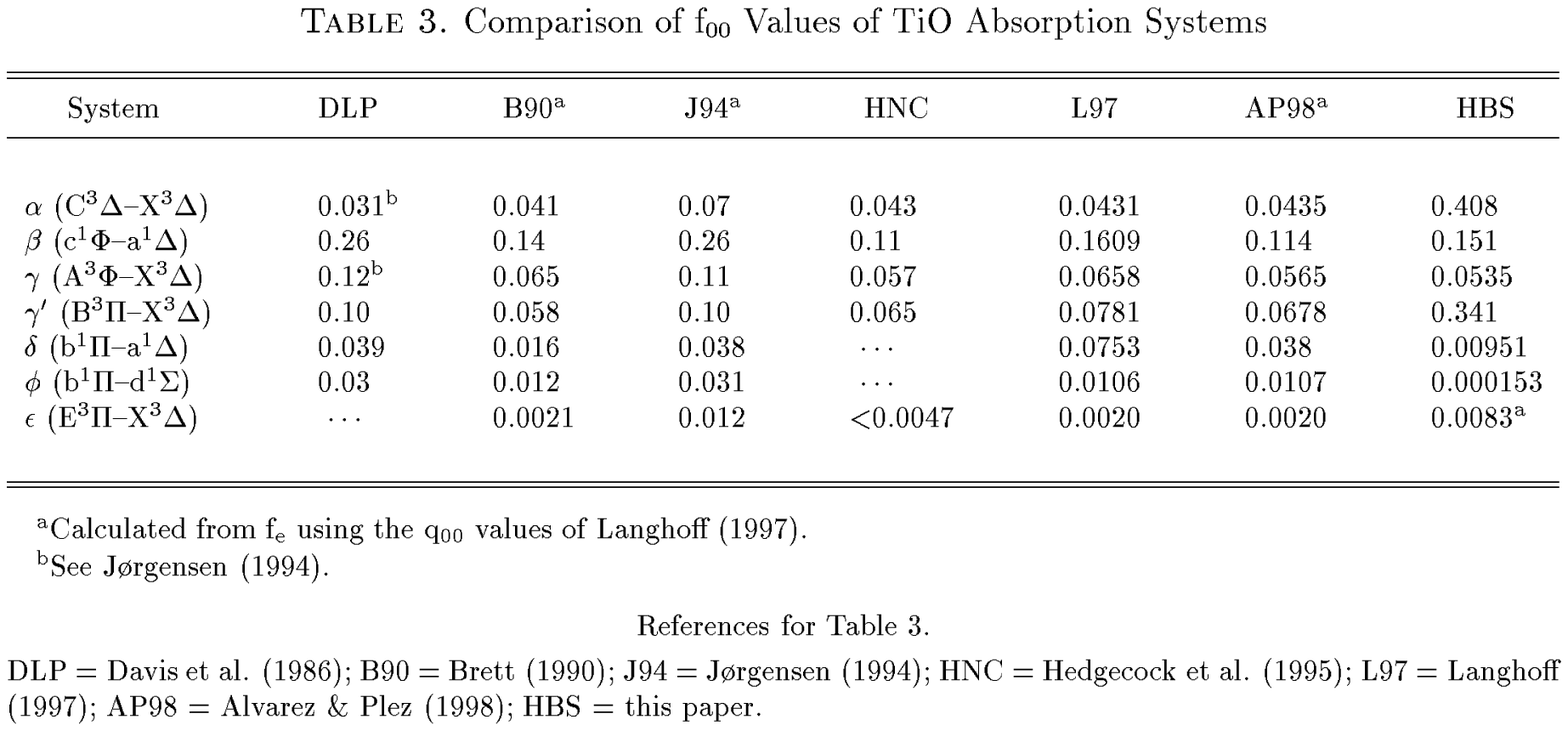}
\end{figure}

\clearpage

\begin{figure}[p]
\vspace*{-1.3in}
\hspace*{-1.0in}
\epsfbox{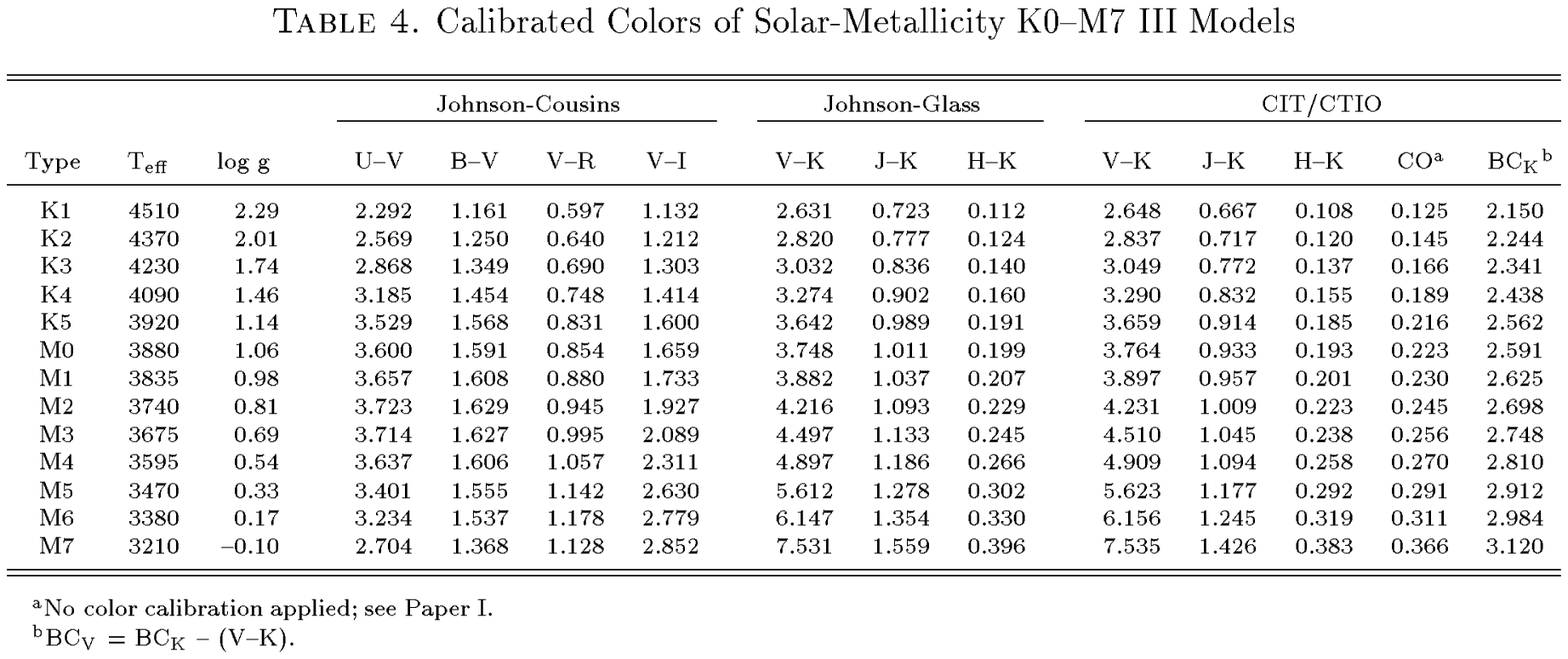}
\end{figure}

\begin{figure}[p]
\vspace*{-1.3in}
\hspace*{-1.0in}
\epsfbox{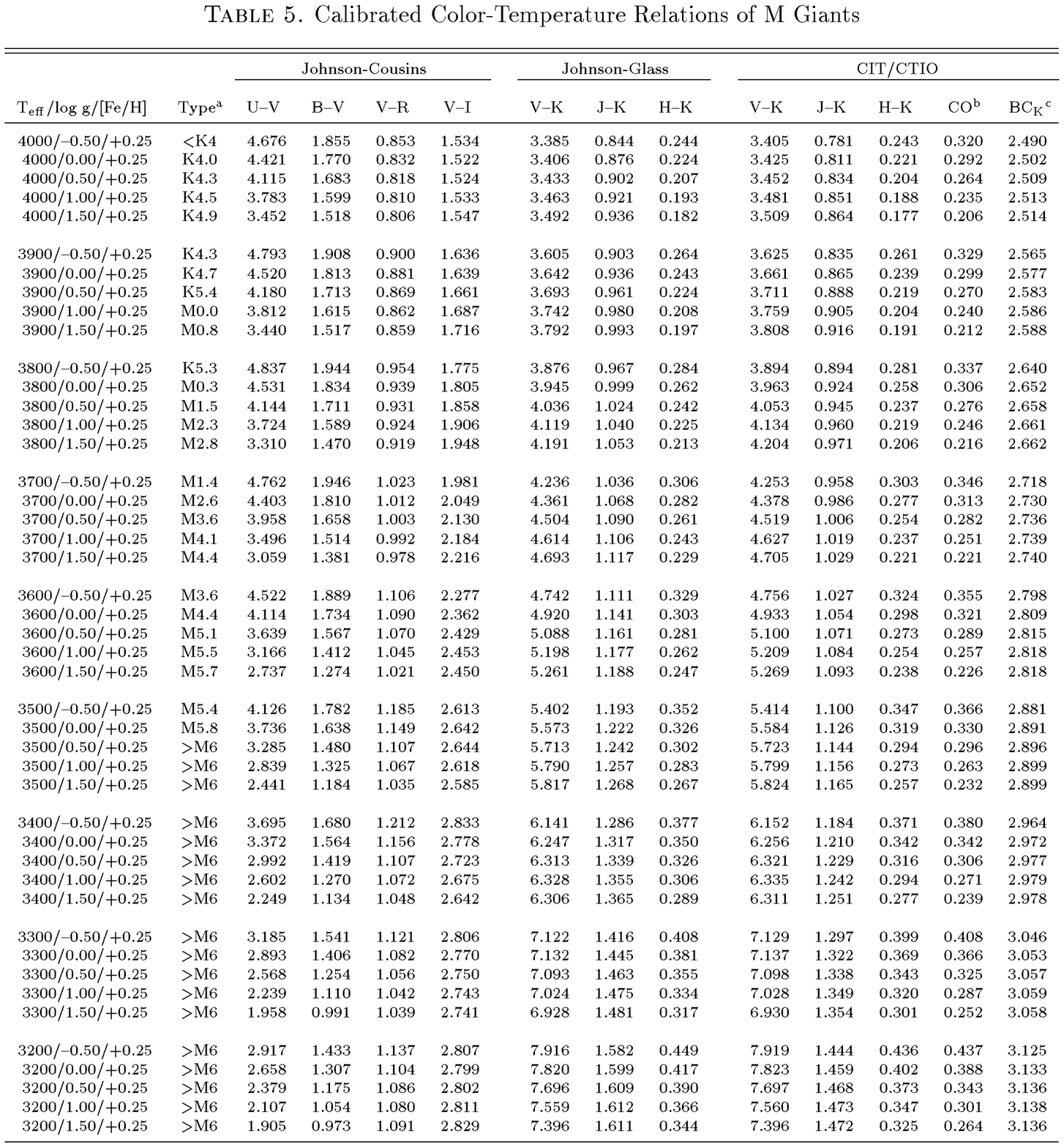}
\end{figure}

\begin{figure}[p]
\vspace*{-1.3in}
\hspace*{-1.0in}
\epsfbox{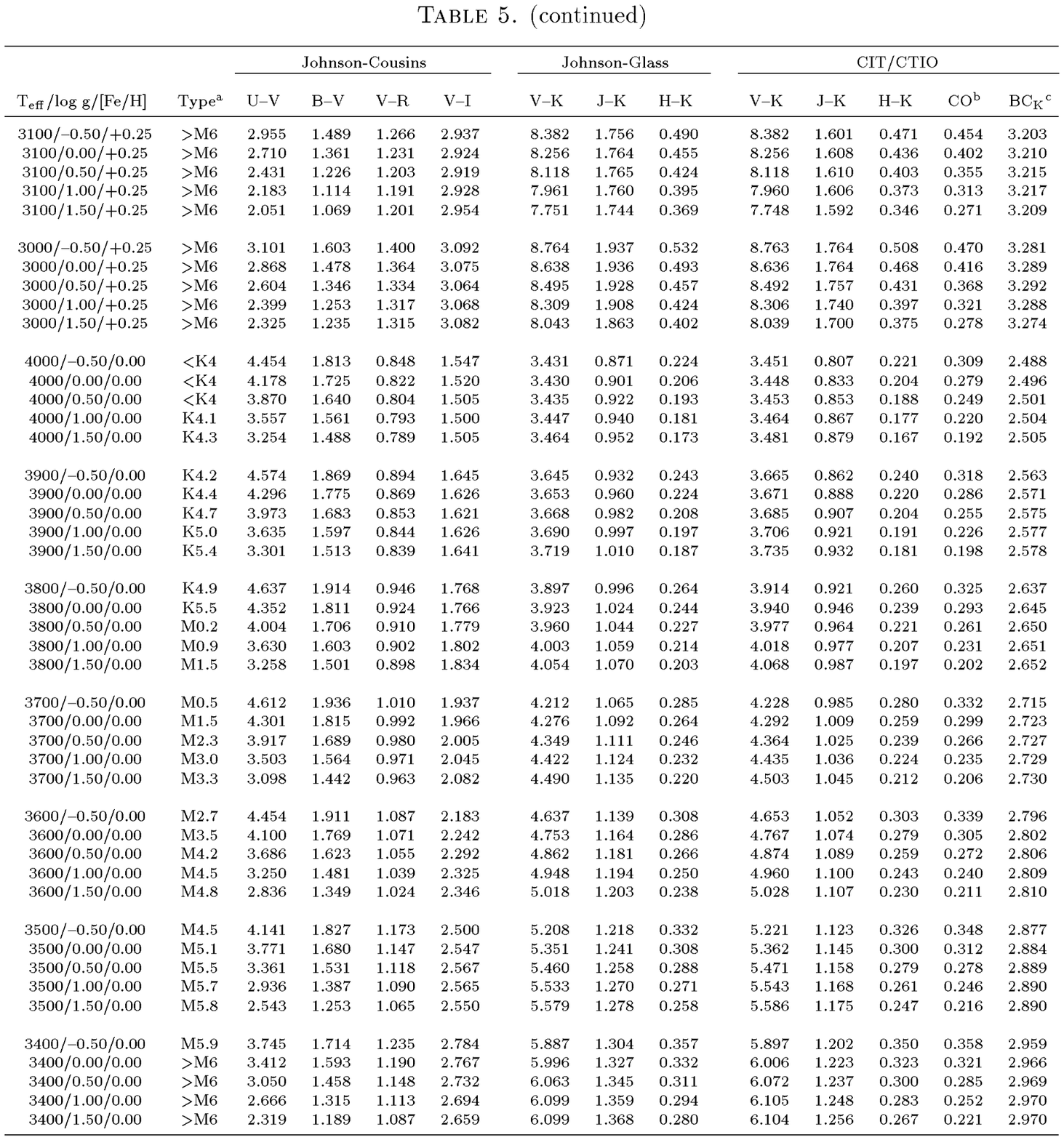}
\end{figure}

\begin{figure}[p]
\vspace*{-1.3in}
\hspace*{-1.0in}
\epsfbox{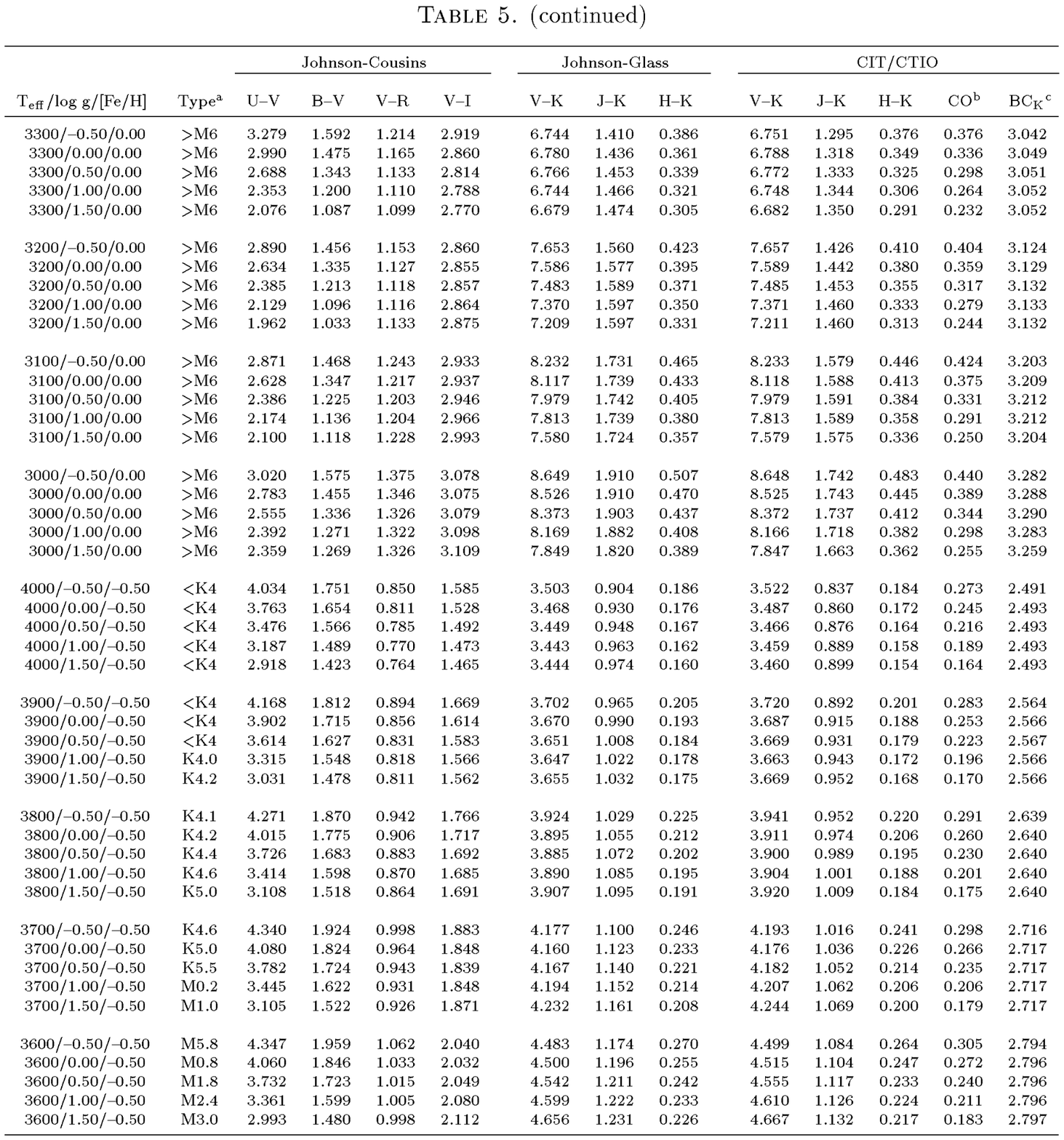}
\end{figure}

\begin{figure}[p]
\vspace*{-1.3in}
\hspace*{-1.0in}
\epsfbox{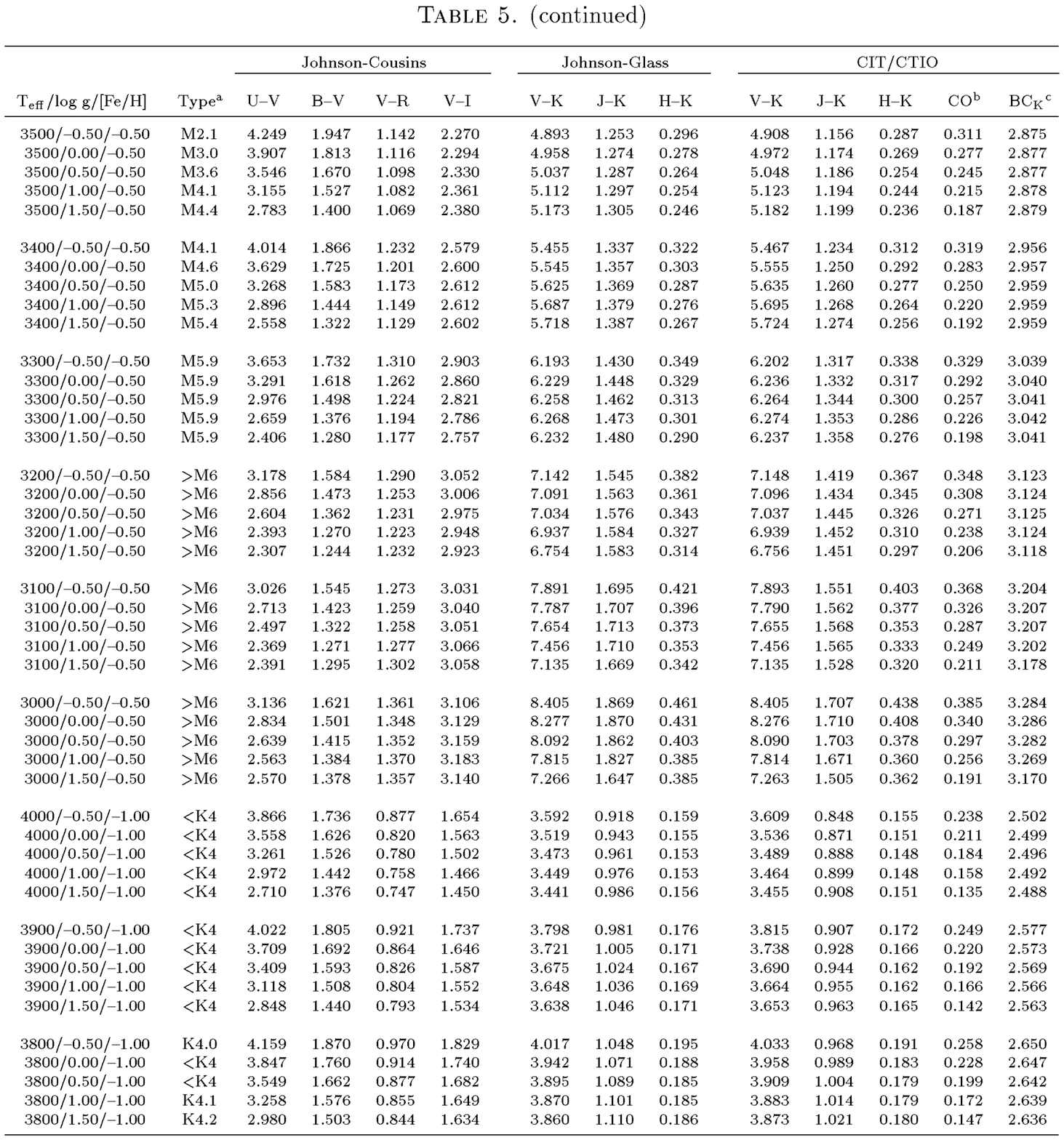}
\end{figure}

\begin{figure}[p]
\vspace*{-1.3in}
\hspace*{-1.0in}
\epsfbox{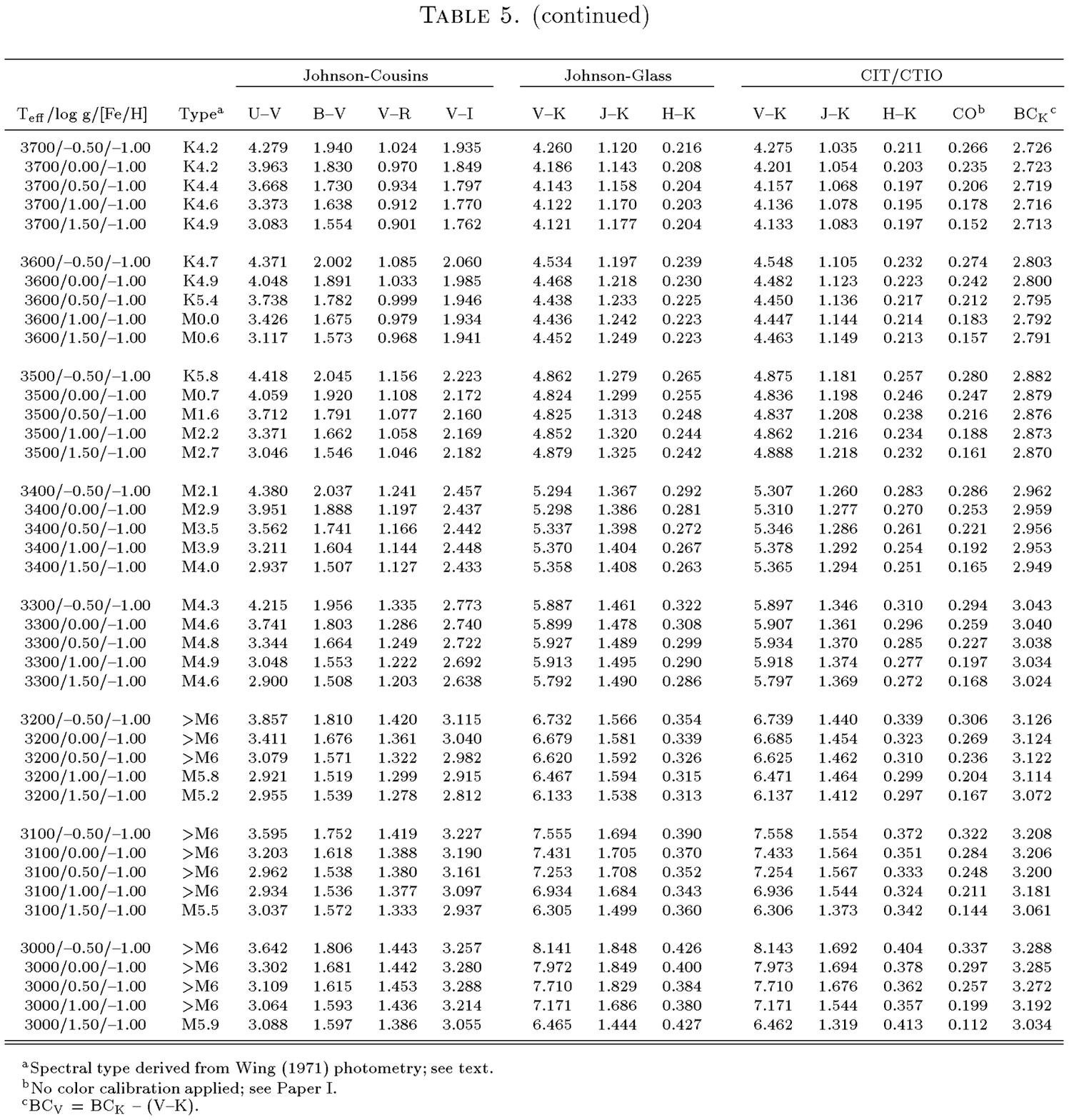}
\end{figure}

\begin{figure}[p]
\vspace*{-1.3in}
\hspace*{-1.0in}
\epsfbox{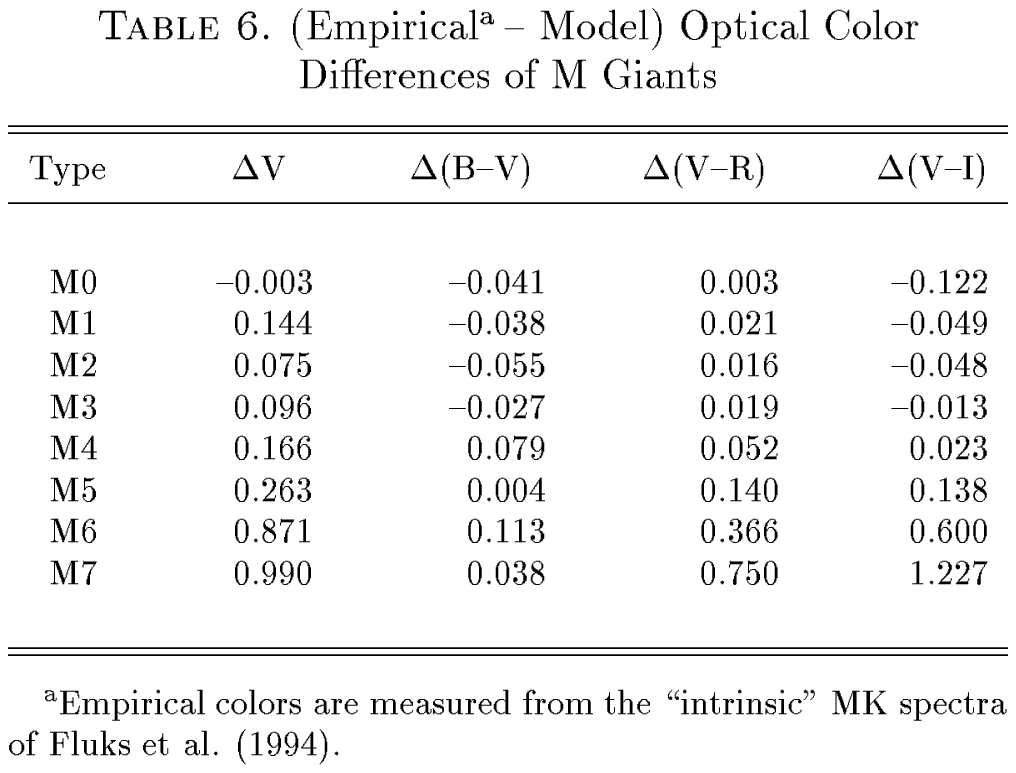}
\end{figure}

\end{document}